\documentclass[conference,a4paper,onecolumn, 11pt]{IEEEtran}  

\usepackage[utf8]{inputenc} 
\usepackage[T1]{fontenc}
\usepackage{url}
\usepackage{ifthen}
\usepackage{cite}
\usepackage[cmex10]{amsmath} 
\usepackage{amsfonts}
\usepackage{amssymb,mathtools}
\usepackage{multirow}
\usepackage{graphicx}
\usepackage{subcaption}    
\usepackage{framed}    

\DeclareMathOperator{\myspan}{span}
\DeclareMathOperator{\col}{col}

\usepackage{tikz}
\usetikzlibrary{tikzmark}
\usetikzlibrary{fit,backgrounds}
\usetikzlibrary{decorations.pathreplacing,angles,quotes}
\usetikzlibrary{patterns}
\usetikzlibrary{calc}
\usetikzlibrary{positioning}
\tikzstyle{block} = [rectangle, draw, 
    minimum width=4em, text centered, rounded corners, minimum height=1em]
\tikzstyle{network} = [rectangle, draw, 
    minimum width=15em, text centered, rounded corners, minimum height=9em]
\tikzstyle{rec} = [rectangle, draw]
\tikzstyle{line} = [draw, -latex, thick]

\usepackage{color}

\newtheorem{lemm}{Lemma}[section]
\newtheorem{defi}{Definition}[section]

\newtheorem{theo}{Theorem}[section]
\newtheorem{prop}{Proposition}[section]

\newcommand{\pp}{\hspace*{0.5em}}

\DeclareMathOperator*{\Tr}{\text{Tr}}

\DeclareMathOperator{\tr}{tr}
\DeclareMathOperator{\cH}{{\mathcal{H}}}
\DeclareMathOperator{\rank}{rank}

\DeclarePairedDelimiter\px{\{}{\}}
\DeclarePairedDelimiter\paren{(}{)}
\DeclarePairedDelimiter\bparen{[}{]}
\DeclarePairedDelimiter\ceil{\lceil}{\rceil}
\DeclarePairedDelimiter\floor{\lfloor}{\rfloor}

\DeclarePairedDelimiter\ket{\lvert}{\rangle}
\DeclarePairedDelimiterX\braket[2]{\langle}{\rangle}{#1 \delimsize\vert #2}
\DeclarePairedDelimiterX\ketbra[2]{\delimsize\vert}{\delimsize\vert}{#1 \rangle\langle #2}

\DeclarePairedDelimiter\bitk{\rvert}{\rangle_{b}}

\DeclarePairedDelimiter\phasek{\lvert}{\rangle_p }

\newcommand\kb[1]{| #1 \rangle\langle #1 |}

\newcommand\zerom[2]{\mathbf{0}_{#1,#2}}
\newcommand\lefto[1]{\mathcal{L}(#1)}
\newcommand\righto[1]{\mathcal{R}(#1)}
\newcommand\leftop[1]{\mathcal{L'}(#1)}
\newcommand\rightop[1]{\mathcal{R'}(#1)}

\newcommand\ph[1]{\widehat{#1}}

\def\xm{\bar{m}}

\def\mm{m_i}
\def\vsec{V_i}
\def\FF{\mathbb{F}}
\def\p{r}
\def\size{1.935}

\begin{document}
\title{
Quantum Network Code for Multiple-Unicast Network with Quantum Invertible Linear Operations
}

\author{%
 \IEEEauthorblockN{Seunghoan Song$^{a}$
 and Masahito Hayashi$^{a,b,c}$}
  \IEEEauthorblockA{$~^{a}$Graduate School of Mathematics, Nagoya University \\
$^b$Centre for Quantum Technologies, National University of Singapore \\
$^c$Shenzhen Institute for Quantum Science and Engineering, Southern University of Science and Technology\\
    Email: \{m17021a, masahito\}@math.nagoya-u.ac.jp} 
}

\maketitle

\begin{abstract}
This paper considers the communication over a quantum multiple-unicast network where $r$ sender-receiver pairs communicate independent quantum states.
We concretely construct a quantum network code for the quantum multiple-unicast network as a generalization of the code [Song and Hayashi, arxiv:1801.03306, 2018] for the quantum unicast network.
When the given node operations are restricted to invertible linear operations between bit basis states
and the rates of transmissions and interferences are restricted,
our code certainly transmits a quantum state for each sender-receiver pair by $n$-use of the network
asymptotically, which guarantees no information leakage to the other users.
Our code is implemented only by the coding operation in the senders and receivers
and employs no classical communication and no manipulation of the node operations.
Several networks that our code can be applied are also given.
\end{abstract}

\section{Introduction} \label{sec:intro}

When we transmit information via network, 
it is useful to make codes by reflecting the network structure.
Such type of coding is called network coding and was initiated by Ahlswede et al. \cite{ACLY}.
This topic has been extensively researched by many researchers.
Network coding employs computation-and-forward in intermediate nodes instead of the naive routing method in traditional network communication.
For the quantum network, the paper \cite{Hayashi2007} started the discussion of the quantum network coding, and many papers \cite{PhysRevA.76.040301, Kobayashi2009, Leung2010, Kobayashi2010,Kobayashi2011} have advanced the study of quantum network coding.

In the network coding, 
unicast network is the most basic network model that the entire network is used by a sender and a receiver.
As one of the remarkable achievements of network coding for the unicast network, on the classical linear network with malicious adversaries,
the papers \cite{Jaggi2008,HOKC17} proposed codes that implement the classical communication by asymptotic $n$-use of the network.
In \cite{Jaggi2008,HOKC17}, when the transmission rate $m$ in absence of attacks is at least the maximum rate $a$ of attack (i.e., $a<m$),
the codes in \cite{Jaggi2008,HOKC17} implement the rate $m-a$ communication asymptotically.
As a quantum generalization of the codes in \cite{Jaggi2008,HOKC17}, 
the paper \cite{SM18} constructed a quantum network code that transmits a quantum state correctly and secretly by asymptotic $n$-use of the network.
Similarly to \cite{Jaggi2008,HOKC17}, when the transmission rate $m$ without attacks is at least twice of the maximum number $a$ of the attacked edges (i.e., $2a < m$), the code in \cite{SM18} implements the rate $m-2a$ quantum communication asymptotically.

However, since a network is used by several users in general, it is needed to treat the network model with multiple users instead of the unicast network.
For this purpose, the multiple-unicast network has been researched, in which disjoint $\p$ sender-receiver pairs $(S_1,T_1), \ldots, (S_\p,T_\p)$ communicate over a network.
The paper \cite{OKH17-2} studied a quantum network code for the multiple-unicast network.
The code in \cite{OKH17-2} transmits a state successfully for each use of the network.
However, \cite{OKH17-2} has a limitation that the code should manipulate the node operations in the network and therefore the code depends on the network structure.
In addition, the code in \cite{OKH17-2} requires the free use of the classical communication.

This paper proposes a quantum network code for the multiple-unicast network 
which is a generalization of the unicast quantum network code in \cite{SM18} 
and overcomes the shortcomings of the multiple-unicast quantum network code in \cite{OKH17-2}.
In the same way as \cite{SM18}, 
the given node operations are invertible linear with respect to the bit basis states, which is called \textit{quantum invertible linear operations} described in Section \ref{sec:network},
our code requires the asymptotic $n$-use of the network for the correct transmission of the state,
and
the encoding and decoding operations are performed on the input and output quantum systems of the $n$-use of the network, respectively.
On the other hand, differently from \cite{OKH17-2}, our code can be implemented without any manipulation of the network operations and any classical communication.
Moreover, our code makes no information leakage asymptotically from a sender $S_i$ to the receivers other than $T_i$ because the correctness of the transmitted state guarantees no information leakage \cite{Schumacher96}.

To discuss the achievable rate by our code, we consider the situation that the input states of all senders are the bit basis states.
Then, our network can be considered as a classical network, called {\em bit classical network}, because 
a bit basis state is transformed to another bit basis state by our quantum node operations.
In the bit classical network,
we assume that 
when the inputs of the senders other than $S_i$ are to zero,
{\em the transmission rate} from $S_i$ to $T_i$ is $m_i$, which is the same as the number of outgoing edges of $S_i$ and incoming edges of $T_i$.
Also, when we define {\em the interference rate} by the rate of the transmitted information to $T_i$ from the senders other than $S_i$,
we assume that the interference rate to $T_i$ is at most $a_i$ in the bit classical network.
In the same way, in case that the input states of all senders are set to the phase basis states (defined in Section \ref{sec:network}),
we call the network as {\em phase classical network}.
In the phase classical network, we also assume that the transmission rate from $S_i$ to $T_i$ is $m_i$ when the inputs of the senders other than $S_i$ are zero.
Also, the interference rate to $T_i$ is at most $a_i'$ in the phase classical network.
Under these constraints, if $a_i+a_i'< m_i$, our code achieves the rate $m_i-a_i-a_i'$ quantum communication from $S_i$ to $T_i$ asymptotically.

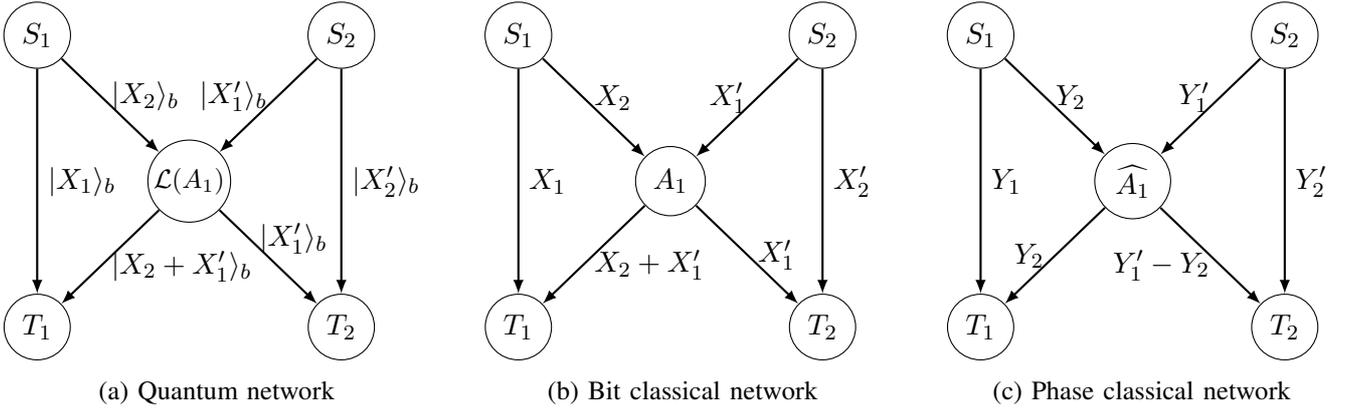
\begin{figure}
    \begin{subfigure}[b]{.333\linewidth}
        \centering
\begin{tikzpicture}[node distance = 3cm, auto]
    \node (x_i) at (0,2.6) {};
    \node [circle,draw](x_1) at (-2,\size) {$S_1$};
    \node [circle,draw](x_5) at (2,\size) {$S_2$};
    \node [circle,draw](y_1) at (-2,-\size) {$T_1$};
    \node [circle,draw](y_5) at (2,-\size) {$T_2$};
    \node [circle,draw,inner sep=1pt] (channel) {\small $\lefto{A_1}$};
    \path [line] (x_1)--node[right,       pos=0.4] {$\bitk{X_2}$} (channel) ;
    \path [line] (x_1)--node {$\bitk{X_1}$} (y_1);
    \path [line] (x_5)--node[left,       pos=0.4] {$\bitk{X'_1}$} (channel);
    \path [line] (x_5)--node {$\bitk{X'_2}$} (y_5);
    \path [line] (channel)--node[right,       pos=0.6] {$\bitk{X_2+X'_1}$}(y_1);
    \path [line] (channel)--node[right,       pos=0.3] {$\bitk{X'_1}$}(y_5);
\end{tikzpicture}
        \caption{Quantum network}
        \label{subfig:q}
    \end{subfigure}
    \begin{subfigure}[b]{.333\linewidth}
        \centering
\begin{tikzpicture}[node distance = 3cm, auto]
    \node (x_i) at (0,2.6) {};
    \node [circle,draw] (channel) {$A_1$};
    \node [circle,draw](x_1) at (-2,\size) {$S_1$};
    \node [circle,draw](x_5) at (2,\size) {$S_2$};
    \node [circle,draw](y_1) at (-2,-\size) {$T_1$};
    \node [circle,draw](y_5) at (2,-\size) {$T_2$};
    \path [line] (x_1)--node[right,       pos=0.4] {$X_2$} (channel) ;
    \path [line] (x_1)--node {$X_1$} (y_1);
    \path [line] (x_5)--node[left,       pos=0.4] {$X'_1$} (channel);
    \path [line] (x_5)--node {$X'_2$} (y_5);
    \path [line] (channel)--node[right,       pos=0.6] {$X_2+X'_1$}(y_1);
    \path [line] (channel)--node[right,       pos=0.5] {$X'_1$}(y_5);
\end{tikzpicture}
        \caption{Bit classical network}
        \label{subfig:b}
    \end{subfigure}%
    \begin{subfigure}[b]{.333\linewidth}
        \centering
\begin{tikzpicture}[node distance = 3cm, auto]
    \node (x_i) at (0,2.6) {};
    \node [circle,draw] (channel) {$\ph{A_1}$};
    \node [circle,draw](x_1) at (-2,\size) {$S_1$};
    \node [circle,draw](x_5) at (2,\size) {$S_2$};
    \node [circle,draw](y_1) at (-2,-\size) {$T_1$};
    \node [circle,draw](y_5) at (2,-\size) {$T_2$};
    \path [line] (x_1)--node[right,       pos=0.4] {$Y_2$} (channel) ;
    \path [line] (x_1)--node {$Y_1$} (y_1);
    \path [line] (x_5)--node[left,       pos=0.4] {$Y'_1$} (channel);
    \path [line] (x_5)--node {$Y'_2$} (y_5);
    \path [line] (channel)--node[left,       pos=0.5] {$Y_2$}(y_1);
    \path [line] (channel)--node[left,       pos=0.6] {$Y'_1-Y_2$}(y_5);
\end{tikzpicture}
        \caption{Phase classical network}
        \label{subfig:p}
    \end{subfigure}
\caption{Toy example of a multiple-unicast network. 
In quantum network (a), $\bitk{\cdot}$ denotes a bit basis state and $\lefto{A_1}$ is the network operation (see Section \ref{sec:network}).
The network (b) and (c) is the bit and phase classical networks of the quantum network (a).} \label{fig:butterfly}
\end{figure}

To help the understanding of the rates described above, we explain the achievable transmission rate from $S_1$ to $T_1$ in the network in Fig. \ref{fig:butterfly}.
The bit and the phase classical networks (Fig. \ref{subfig:b} and Fig. \ref{subfig:p}) are determined from the quantum network (Fig. \ref{subfig:q}) (see Section \ref{sec:network}).
When $X'_1=X'_2=Y'_1=Y'_2=0$, the transmission rates from $S_1$ to $T_1$ are $2$ for both networks, i.e., $m_1=2$, which is also the number of outgoing edges of $S_1$ and incoming edges of $T_1$. 
Also, the interference rates from $S_2$ to $T_1$ are $1$ and $0$ for the bit and the phase classical networks, respectively.
On this network, 
if our code from $S_1$ to $T_1$ with the rates $(m_1,a_1,a_1')=(2,1,0)$ is constructed,
the conditions $a_1\geq 1$, $a_1'\geq0$ and $a_1+a_1'<m_1$ are satisfied, and therefore our code implements the rate $m_1-a_1-a_1'=1$ quantum transmission from $S_1$ to $T_1$ asymptotically.

In the practical sense, our code can cope with the node malfunctions in the following case:
on the multiple-unicast network with quantum invertible linear operations,
the network operations 
are well-determined so that there is no interference between all sender-receiver pairs, 
but three broken nodes apply quantum invertible linear operations different from the determined ones.
Moreover, let the transmission rate $m_1$ without interferences from $S_1$ to $T_1$ be $100$ and the number of outgoing edges of the three broken nodes be $4$.
In this case, $3\times 4=12$ outgoing edges of the three broken nodes transmit the unexpected information which implies the bit (phase) interference rate is at most $12$.
Therefore, by our code with $m_1=100$ and $a_1,a_1'>12$, the sender $S_1$ can transmit quantum states to the receiver $T_1$ correctly with the rate $100-a_1-a_1' < 76$ by asymptotically many uses of the network.

The remaining of this paper is organized as follows.
Section \ref{sec:network} introduces the formal description of the quantum multiple-unicast network with quantum invertible linear operations.
Section \ref{sec:main_result} gives the main results of this paper. 
Based on the preliminaries in Section \ref{sec:prelim}, 
Section \ref{sec:code} concretely constructs our code with the free use of negligible rate shared randomness. 
The encoder and decoder of our code is given in this section.
Section \ref{sec:analysis} analyzes the correctness of the code in Section \ref{sec:code}.
Then, 
Section \ref{sec:without_sr} constructs our code without the assumption of shared randomness 
by attaching the secret and correctable communication protocol \cite{YSJL14} 
to the code given in Section \ref{sec:code},
which proves the main result given in Section \ref{sec:main_result}.
Section \ref{sec:network_examples} gives several examples of the network that our code can be applied.
Section \ref{sec:conclusion} is the conclusion of this paper.

\section{Quantum Network with Invertible Linear Operations} \label{sec:network}
Our code is designed on the quantum network which is a generalization of a classical multiple-unicast network.
In this section, we first introduce the multiple-unicast network with classical invertible linear operations and 
generalize this network as a network with quantum invertible linear operations.
The node operations introduced in this section are identical to the operations in \cite[Section II]{SM18}.

\subsection{Classical Network with Invertible Linear Operations}    \label{subsec:network_classic}

First, we describe the multiple-unicast network with classical invertible linear operations.
The network topology is given as a directed Graph $G=(V,E)$.
The $\p$ senders and $\p$ receivers are given as $\p$ source nodes $S_1,\ldots, S_\p$ and $\p$ terminal nodes $T_1,\ldots,T_\p$.
The sender $S_i$ has $m_i$ outgoing edges and the receiver $T_i$ has $m_i$ incoming edges.
Define $m:= m_1+\cdots+m_\p$.
The intermediate nodes are numbered from $1$ to $c$ ($=|V|-2\p$) accordingly to the order of the transmission.
The intermediate node numbered $t$ has the same number $k_t$ of incoming and outgoing edges where $1\leq k_t \leq m$.

Next, we describe the transmission and the operations on this network.
Each edge sends an element of the finite field $\mathbb{F}_q$ where $q$ is a power of a prime number $p$.
The $t$-th node operation is described as an invertible linear operation $A_t$ from the information on $k_t$ incoming edges to that of $k_t$ outgoing edges.
Since node operations are invertible linear,
the entire network operation is written as $K = A_{c}\cdots A_{1} \in \mathbb{F}_q^{m\times m}$.
For the network operation $K$, we introduce the following notation:
\begin{align*}
K := 
\begin{bmatrix}
 K_{1,1} & K_{1,2} & \cdots &  K_{1,\p}\\
 K_{2,1} & K_{2,2} & \cdots &  K_{2,\p}\\
 \vdots  &  \ddots &        &  \vdots \\
 K_{\p,1} & K_{\p,2} & \cdots &  K_{\p,\p}\\
 \end{bmatrix}
 ,\quad
K_{i,j} \in \mathbb{F}_q^{m_i\times m_j}.
\end{align*}
Then, $K_{i,j}$ is the network operation from $S_i$ to $T_j$. 
We assume $\rank K_{i,i} = m_i$ which means the information from $S_i$ to $T_i$ is completely transmitted if there is no interference.

When the network inputs by senders $S_1,\ldots, S_\p$ are $x_1\in\mathbb{F}_q^{m_1},\ldots, x_\p\in\mathbb{F}_q^{m_\p}$, 
the output $y_i\in\mathbb{F}_q^{m_i}$ at the receiver $T_i$ ($i=1,\ldots,\p$) is written as 
\begin{align}
y_i =& \sum_{j=1}^{\p} K_{i,j} x_j 
                                  = K_{i,i}x_i + K_{i^c} z_{i^c}  , \label{def:interf}\\
K_{i^c} :=& [K_{i,1}\pp \cdots \pp K_{i,i-1} \pp K_{i,i+1} \pp \cdots \pp K_{i,\p}]\in\mathbb{F}_q^{m_i\times (m-m_i)}, \nonumber\\
z_{i^c} :=& [x_1^{\mathrm{T}} \pp \cdots \pp x_{i-1}^{\mathrm{T}} \pp x_{i+1}^{\mathrm{T}} \pp \cdots \pp x_\p^{\mathrm{T}}]^{\mathrm{T}}\in\mathbb{F}_q^{m-m_i}. \nonumber
\end{align}
The second term $K_{i^c} z_{i^c}$ of \eqref{def:interf} is called the interference to $T_i$,
and $\rank K_{i^c}$ is called the rate of the interference to $T_i$.

Consider the $n$-use of the above network.
When the inputs by senders $S_1,...,S_\p$ are $X_1\in\mathbb{F}_q^{m_1\times n},\ldots, X_\p\in\mathbb{F}_q^{m_\p\times n}$, the output $Y_i\in\mathbb{F}_q^{m_i\times n}$ at the receiver $T_i$ ($i=1,\ldots,\p$) is 
\begin{align*}
Y_i =& \sum_{j=1}^{\p} K_{i,j} X_j 
    = K_{i,i}X_i + K_{i^c} Z_{i^c},\\
Z_{i^c} :=& [X_1^{\mathrm{T}} \pp \cdots \pp X_{i-1}^{\mathrm{T}} \pp X_{i+1}^{\mathrm{T}} \pp \cdots \pp X_\p^{\mathrm{T}}]^{\mathrm{T}} \in \mathbb{F}_q^{(m-m_i)\times n}.
\end{align*}


\subsection{Quantum Network with Invertible Linear Operations}
We generalize the multiple-unicast network with classical invertible linear operations to the network with quantum invertible linear operations.
In this quantum network, the network topology is the same graph $G=(V,E)$.
Each edge transmits a quantum system $\cH$ which is $q$-dimensional Hilbert space spanned by the bit basis $\{\bitk{x}\}_{x\in\mathbb{F}_q}$.
In $n$-use of the network, 
we treat the quantum system $\cH^{\otimes m_i\times n}$ spanned by the bit basis $\{\bitk{X}\}_{X\in\mathbb{F}_q^{m_i\times n}}$.
The sender $S_i$ sends a quantum system $\cH_{S_i} := \cH^{\otimes m_i\times n}$ and the receiver $T_i$ receives a quantum system $\cH_{T_i}:= \cH^{\otimes m_i\times n}$

To describe the quantum node operation, we define the following quantum operations.
\begin{defi}[Quantum Invertible Linear Operation] \label{def:q_lin_op}
For invertible matrices $A \in \mathbb{F}_q^{m\times m}$ and $B \in \mathbb{F}_q^{n\times n}$,
two unitaries $\lefto{A}$ and $\righto{B}$ are defined for any $X\in\mathbb{F}_q^{m\times n}$ as
\begin{align*}
\lefto{A} \bitk{X} &:= \bitk{A X}, \quad
\righto{B} \bitk{X} := \bitk{XB}. \nonumber
\end{align*}
The operations $\lefto{A}$ and $\righto{B}$ are called {\it quantum invertible linear operations}.
\end{defi}

The $t$-th node operation is given as $\lefto{A_t}$ and it is called quantum invertible linear operation.
The entire network operation is written as the unitary $\lefto{K} = \lefto{A_{c}\cdots A_{1}} = \lefto{A_{c}}\cdots \lefto{A_{1}}$.
When a state $\rho$ on $\cH_{S_1}\otimes \cdots \otimes \cH_{S_\p}$ is transmitted by senders $S_1,\ldots,S_{\p}$, the network output $\sigma_{T_i}$ at $\cH_{T_i}$ is written as 
    \begin{align*}
    \sigma_{T_i} := \Tr_{T_1,\ldots,T_{i-1},T_{i+1},\ldots,T_{\p}} \lefto{K}\rho\lefto{K}^{\dagger},
    \end{align*}
    where $\Tr_{T_1,\ldots,T_{i-1},T_{i+1},\ldots,T_{\p}}$ is the partial trace on the system $\cH_{T_1}\otimes \ldots \otimes \cH_{T_{i-1}} \otimes \cH_{T_{i+1}} \otimes \ldots \otimes \cH_{T_{\p}}$.

When the input state on the network is $\bitk*{M}$ on $\cH_{S_1}\otimes \cdots \otimes \cH_{S_\p}$, this quantum network can be considered as the classical network in Subsection \ref{subsec:network_classic}.
In the same way as the classical network, we assume $\rank K_{i,i} = m_i$
which means $S_i$ transmits any bit basis states completely to $T_i$ if the input states on source nodes $S_j$ ($j\neq i$) are zero bit basis states.
Similarly, $\rank K_{i^c}$ is called the rate of the bit interference to $T_i$.

We can discuss the interference similarly on the phase basis $\{ \phasek{z} \}_{z\in\mathbb{F}_q}$ defined in \cite[Section 8.1.2]{Haya2} by 
\begin{align*}
\phasek{z} := \frac{1}{\sqrt{q}} \sum_{x\in\mathbb{F}_q} \omega^{-\tr xz} \bitk{x},
\end{align*}
where $\omega := \exp{\frac{2\pi i}{p}}$ and $\tr y:= \Tr M_y$ ($y\in\mathbb{F}_q$) with
the multiplication map $M_y:x \mapsto yx$ identifying the finite field $\mathbb{F}_q$ with the vector space $\mathbb{F}_p^t$.
For the analysis of the phase basis interference, we give Lemma \ref{lemm:invertible_to_unitary} which explains how node operations $\lefto{A_t}$ are applied to the phase basis states. 
\begin{lemm}[{\cite[Appendix A]{SM18}}]\label{lemm:invertible_to_unitary}
Let $A \in \mathbb{F}_q^{m\times m}$ and $B \in \mathbb{F}_q^{n\times n}$ be invertible matrices.
For any $M\in\mathbb{F}_q^{m\times n}$, we have
\begin{align*}
\lefto{A} \phasek{M} = \phasek{(A^{\mathrm{T}})^{-1}M},\quad
\righto{B}\phasek{M} = \phasek{M(B^{\mathrm{T}})^{-1}}.
\end{align*}
\end{lemm}
For notational convenience, we denote $\ph{A} := (A^{\mathrm{T}})^{-1}$.
When the input state is a phase basis state $\phasek{M}$ on $\cH_{S_1}\otimes \cdots \otimes \cH_{S_\p}$, the network operation $\lefto{K}$ is applied by $\lefto{K} \phasek{M} = \phasek{\ph{K}M}$.
In this case, this quantum network can also be considered as a classical network with network operation $\ph{K} = \ph{A_c}\cdots\ph{A_1}$.
Then, $\ph{K}_{i,j}$ is defined from $\ph{K}$ in the same way as $K_{i,j}$.
    \begin{align*}
\ph{K} :=&
\begin{bmatrix}
 \ph{K}_{1,1} & \ph{K}_{1,2} & \cdots &  \ph{K}_{1,\p}\\
 \ph{K}_{2,1} & \ph{K}_{2,2} & \cdots &  \ph{K}_{2,\p}\\
 \vdots  &  \ddots &        &  \vdots \\
 \ph{K}_{\p,1} & \ph{K}_{\p,2} & \cdots &  \ph{K}_{\p,\p}\\
 \end{bmatrix}
 ,\quad
\ph{K}_{i,j} \in \mathbb{F}_q^{m_i\times m_j}, \\
\ph{K}_{i^c} :=& [\ph{K}_{i,1}\pp \cdots \pp \ph{K}_{i,i-1} \pp \ph{K}_{i,i+1} \pp \cdots \pp \ph{K}_{i,\p}].
    \end{align*}
Similarly to the condition $\rank K_{i,i}=m_i$, we also assume $\rank \ph{K}_{i,i} = m_i$.
We also call $\rank \ph{K}_{i^c}$ the rate of phase interference to $T_i$.
The transmission rates from $S_i$ to $T_i$ are summarized in Table \ref{tab:rates}.
    



\renewcommand{\arraystretch}{1.02}
\begin{table}
    \caption{Definitions of Information Rates} \label{tab:rates}
    \begin{center}
    \begin{tabular}[ht]{|c|c|}
    \hline
    Rate & Meaning  \\
    \hline 
    & \\[-1.1em]
    $m_i=\rank K_{i,i}=\rank \ph{K}_{i,i}$ & Bit (phase) transmission rates from $S_i$ to $T_i$ without interference\\
    $\rank K_{i^c}$ & Rate of interference to $T_i$ \\ 
    $\rank \ph{K}_{i^c}$ & Rate of phase interference to $T_i$ \\ 
    $a_i$ & Maximum rate of bit interference to $T_i$\\ 
    $a_i'$ & Maximum rate of phase interference to $T_i$\\ \hline
    \end{tabular}
    \end{center}
\end{table}
\renewcommand{\arraystretch}{1}

\section{Main Results}  \label{sec:main_result}

In this section, we propose two main theorems of this paper.
The two theorems state the existence of our code with and without negligible rate shared randomness, respectively.
The codes stated in the theorems are concretely constructed in Section \ref{sec:code} and \ref{sec:without_sr}, respectively.
The theorems are stated with respect to the completely mixed state $\rho_{mix}$ 
and the {\it entanglement fidelity} $F_e^2(\rho,\kappa):=\langle x |\kappa\otimes\iota_R(\kb{x})|x\rangle$ for the quantum channel $\kappa$ and a purification $|x\rangle$ of the state $\rho$.

\begin{theo} \label{theo:main1}
Consider the transmission from the sender $S_i$ to the receiver $T_i$ over a quantum multiple-unicast network with quantum invertible linear operations given in Section \ref{sec:network}.
Let 
$m_i$ be the bit and phase transmission rates from $S_i$ to $T_i$ without interferences ($m_i=\rank K_{i,i}=\rank \ph{K}_{i,i}$),
and 
$a_i, a_i'$ be the upper bounds of the bit and phase interferences, respectively ($\rank K_{i^c} \leq a_i $, $\rank \ph{K}_{i^c}\leq a'_i $).
When 
the condition $a_i+a'_i < m_i$ holds and
the sender $S_i$ and receiver $T_i$ can share the randomness whose rate is negligible in comparison with the block-length $n$,
there exists a quantum network code whose rate is $m_i-a_i-a_i'$ and the entanglement fidelity $F_e^2(\rho_{mix},\kappa_i)$ satisfies $n(1-F_e^2(\rho_{mix},\kappa_i))\to 0$ where $\kappa_i$ is the quantum code protocol from sender $S_i$ to receiver $T_i$.
\end{theo}

Section \ref{sec:code} constructs the code stated in Theorem \ref{theo:main1} and Section \ref{sec:analysis} shows that this code has the performance in Theorem \ref{theo:main1}.
Note that this code does not depend on the detailed network structure, but depends only on the information rates $m_i,a_i$ and $a'_i$.
As explained in \cite[Section III]{SM18}, 
our code has no information leakage 
from the condition $n(1-F_e^2(\rho_{mix},\kappa_i))\to 0$.

Although Theorem \ref{theo:main1} assumed the free use of the negligible rate shared randomness,
it is possible to design the code of the same performance without negligible rate shared randomness as follows.
The paper \cite{YSJL14} gives the secret and correctable classical network communication protocol for the classical network with malicious attacks,
when the transmission rate is more than the sum of the rate of attacks and the rate of information leakage.
By applying the protocol in \cite{YSJL14} to our quantum network with bit basis states,
the negligible rate shared randomness can be generated.
By this method, we have the following Theorem \ref{theo:main2} and the details are explained in Section \ref{sec:without_sr}.

\begin{theo} \label{theo:main2}
Consider the transmission from the sender $S_i$ to the receiver $T_i$ over a quantum multiple-unicast network with quantum invertible linear operations given in Section \ref{sec:network}.
Let 
$m_i$ be the bit and phase transmission rates from $S_i$ to $T_i$ without interferences ($m_i=\rank K_{i,i}=\rank \ph{K}_{i,i}$),
and 
$a_i, a_i'$ be the upper bounds of the bit and phase interferences, respectively ($\rank K_{i^c} \leq a_i$, $\rank \ph{K}_{i^c}\leq a'_i$).
When $a_i+a'_i < m_i$, there exists a quantum network code whose rate is $m_i-a_i-a_i'$ and the entanglement fidelity $F_e^2(\rho_{mix},\kappa_i)$ satisfies $n(1-F_e^2(\rho_{mix},\kappa_i))\to 0$ where $\kappa_i$ is the quantum code protocol from sender $S_i$ to receiver $T_i$.
\end{theo}

\begin{figure*}[tb]
\begin{center}
\begin{tikzpicture}[every text node part/.style={align=center},node distance = 5.4cm, auto]
    \node [network] (channel) {Quantum Network \\ Multiple-Unicast };
    \node [block] (encoder) at (-6.4,0.4) {Encoder $\mathcal{E}_i^{SR_i,R_i}$};
    \node (pr) at (-6.4,3) {\small (Private Randomness $U_{i,1}$)};
    \node [block] (decoder) at (5,0.4) {Decoder $\mathcal{D}_i^{SR_i}$};
    \node at (-8.7, 0.4) {$\rho_{i}$} ;
    \node at (8.2, 0.4) {$\mathcal{D}_i^{SR_i}(\sigma_{T_i})$} ;
    
    \path [line,dashed] (encoder) edge [<->,bend left=39] node {\small (Shared Randomness $SR_i$)} (decoder) ;
    \path [line] (-8.5,0.4) -- (encoder);
    \path [line] (encoder) -- node {$\mathcal{E}_i^{SR_i,R_i}(\rho_i)$}(-2.8,0.4);
    \path [line] (2.78,0.4) -- node {$\sigma_{T_i}$}(decoder);
    \path [line] (decoder) -- (7.3,0.4);
    \path [line,dashed] (pr) edge [->] (encoder);
    
    \node at (-2.5, 1.4) {\scriptsize \fbox{$S_1$}};
    \node at (-2.5, 1) {\small \vdots};
    \node at (-2.5, 0.4) {\scriptsize \fbox{$S_i$}};
    \node at (-2.5, -0.4) {\small \vdots};
    \node at (-2.5, -1.4) {\scriptsize \fbox{$S_r$}};
    
    \node at (2.5, 1.4) {\scriptsize \fbox{$T_1$}};
    \node at (2.5, 1) {\small \vdots};
    \node at (2.5, 0.4) {\scriptsize \fbox{$T_i$}};
    \node at (2.5, -0.4) {\small \vdots};
    \node at (2.5, -1.4) {\scriptsize \fbox{$T_r$}};
    
    \path [line] (2.8, 1.4) -- (8, 1.4);
    \path [line] (-8, -1.4) -- (-2.8, -1.4);
    \path [line] (2.8, -1.4) -- (8, -1.4);
    \path [line] (-8, 1.4) -- (-2.8, 1.4);
    
\end{tikzpicture}
        \caption{Overview of code protocol from sender $S_i$ to receiver $T_i$.
        States $\rho_{i}$ and $\mathcal{D}_i^{SR_i}(\sigma_{T_i})$ are in code space $\cH'_\text{code}$.
        } \label{fig1}
\end{center}
\end{figure*}
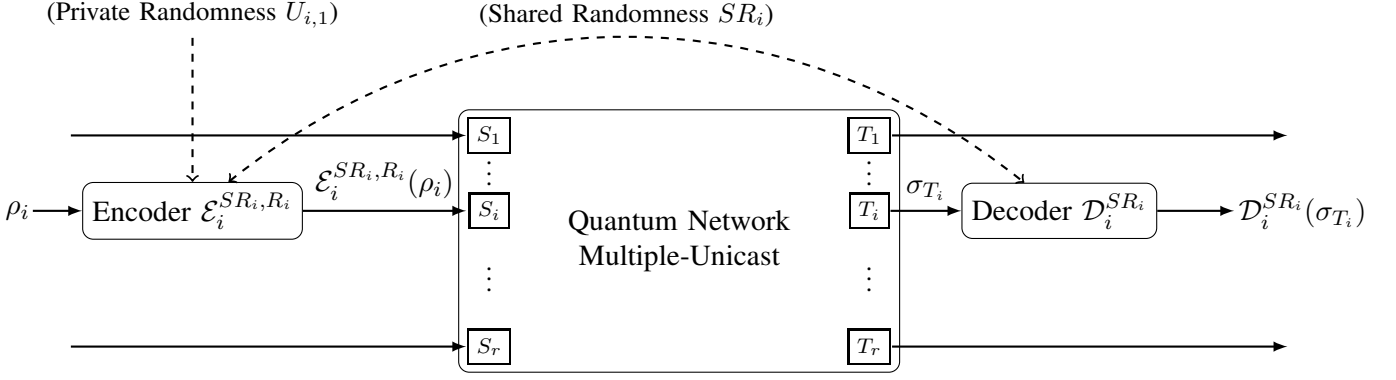

\section{Preliminaries for Code Construction} \label{sec:prelim}
Before code construction, we prepare the extended quantum system, notations, and CSS code used in our code.

\subsection{Extended Quantum System}
Although the unit quantum system for the network transmission is $\cH$, our code is constructed based on the extended quantum system $\cH'$ described below.

First, dependently on the block-length $n$, we choose a power $q':=q^\alpha$ to satisfy $n\cdot (n')^{m_i}/(q')^{m_i-\max\{a_i,a_i\}}\to 0$ (e.g. $q'=O(n^{1+(\max\{a_i,a_i'\}+2)/(m_i-\max\{a_i,a_i'\}))}$ 
) where $n':=n/\alpha$.
Let $\mathbb{F}_{q'}$ be the $\alpha$-dimensional field extension of $\mathbb{F}_q$.
Similarly, let $\cH':=\cH^{\otimes \alpha}$ be the quantum system spanned by $\{\bitk{x}\}_{x \in \FF_{q'}}$.
Then, the $n$-use of the network over $\cH$ can be considered as the $n'$-use of the network over $\cH'$.
The quantum invertible linear operations (Definition \ref{def:q_lin_op}) can also be defined for invertible matrices $A' \in \mathbb{F}_{q'}^{m\times m}$ and $B' \in \mathbb{F}_{q'}^{n\times n}$ as
\begin{align}
\leftop{A} \bitk{X} = \bitk{A X}, \pp
\rightop{B} \bitk{X} = \bitk{XB},  \quad  \text{for any } X\in\mathbb{F}_{q'}^{m\times n} \nonumber.
\end{align}

%

\subsection{Notations for Quantum Systems and States in Our Code} \label{sec:notations}

We introduce notations used in our code.
By the $n$-use of the network,
the sender $S_i$ transmits the system $\cH_{S_i} = \cH^{\otimes m_i\times n}$ and
the receiver $T_i$ receives the system $\cH_{T_i} = \cH^{\otimes m_i\times n}$, which are identical to $\cH'^{\otimes m_i \times n'}$.
We partition the quantum system $\cH'^{\otimes m_i \times n'}$ as $\cH'_{\mathcal{A}} \otimes \cH'_{\mathcal{B}} \otimes \cH'_{\mathcal{C}} := \cH'^{\otimes m_i \times m_i} \otimes \cH'^{\otimes m_i \times m_i} \otimes \cH'^{\otimes m_i \times (n'-2m_i)}$.
Furthermore, we partition the systems $\cH'_{\mathcal{A}}, \cH'_{\mathcal{B}}, \cH'_{\mathcal{C}}$ by 
\begin{align*}
\cH'_{\mathcal{A}} &= \cH'_{\mathcal{A}1} \otimes \cH'_{\mathcal{A}2} \otimes \cH'_{\mathcal{A}3} := \cH'^{\otimes a_i\times m_i} \otimes \cH'^{\otimes (m_i-a_i-a'_i)\times m_i} \otimes \cH'^{\otimes a'_i\times m_i},\\
\cH'_{\mathcal{B}} &= \cH'_{\mathcal{B}1} \otimes \cH'_{\mathcal{B}2} \otimes \cH'_{\mathcal{B}3} := \cH'^{\otimes a_i\times m_i} \otimes \cH'^{\otimes (m_i-a_i-a'_i)\times m_i} \otimes \cH'^{\otimes a'_i\times m_i},\\
\cH'_{\mathcal{C}} &= \cH'_{\mathcal{C}1}\otimes \cH'_{\mathcal{C}2}\otimes \cH'_{\mathcal{C}3} := \cH'^{\otimes a_i\times (n'-2m_i)} \otimes \cH'^{\otimes (m_i-a_i-a'_i)\times (n'-2m_i)} \otimes \cH'^{\otimes a'_i\times (n'-2m_i)}.
\end{align*}

For states $|\phi\rangle\in \cH'_{\mathcal{A}1}, 
                |\psi\rangle\in \cH'_{\mathcal{A}2}$, and 
                $|\varphi\rangle\in \cH'_{\mathcal{A}3}$,
the tensor product state in $\cH'_{\mathcal{A}}$ is denoted as
\begin{align}
    \left[\begin{array}{c}
    \ket*{\phi} \\ 
    |\psi\rangle \\ 
    |\varphi\rangle \\
    \end{array} \right]
    :=
    |\phi\rangle 
    \otimes |\psi\rangle 
    \otimes |\varphi\rangle
    \in
    \cH'_{\mathcal{A}}.
    \label{nota1}
\end{align}
The bit or phase basis state of $(X, Y, Z ) \in \mathbb{F}_{q'}^{a_i\times m_i}\times \mathbb{F}_{q'}^{(m_i-a_i-a'_i)\times m_i}\times \mathbb{F}_{q'}^{a'_i\times m_i}$ is  denoted as
\begin{align}
  \bitk*{\left[\!\!\begin{array}{c}
    X \\ 
    Y \\ 
    Z \\
    \end{array} \!\!\right]
    }
    := 
    \left[\!\!\begin{array}{c}
    |X\rangle_b \\ 
    |Y\rangle_b \\ 
    |Z\rangle_b \\
    \end{array} \!\!\right]
    ,
    \quad
    \phasek*{\left[\!\!\begin{array}{c}
    X \\ 
    Y \\ 
    Z \\
    \end{array} \!\!\right]
    }
    := 
    \left[\!\!\begin{array}{c}
    |X\rangle_p \\ 
    |Y\rangle_p \\ 
    |Z\rangle_p \\
    \end{array} \!\!\right]
    .
    \label{nota2}
\end{align}
We also introduce notations for the states in $\cH'_{\mathcal{B}}$ and $\cH'_{\mathcal{C}}$ in the same way as \eqref{nota1} and \eqref{nota2}.
In the following, we denote the $k\times l$ zero matrix as $\zerom{k}{l}$.


\subsection{CSS Code in Our Code} \label{sec:css_code}

In our code construction, we use the CSS code defined in this subsection which is similarly defined from \cite[Subsection IV-B]{SM18}.
Define two classical codes $C_1, C_2\subset \mathbb{F}_{q'}^{m_i\times (n'-2\mm)}$ which satisfy $C_1 \supset C_2^{\perp}$ as 
\begin{align*}
C_1 :=& \px*{
    \begin{bmatrix}
    \zerom{a_i}{n'-2\mm} \\ 
    X_2 \\ 
    X_3 
    \end{bmatrix}
    \in \mathbb{F}_{q'}^{m_i\times(n'-2\mm)}
    \bigg|
    X_2\in \mathbb{F}_{q'}^{(m_i-a_i-a_i')\times(n'-2\mm)},
           X_3\in \mathbb{F}_{q'}^{a_i'\times(n'-2\mm)}
    },\\
C_2 :=& \px*{
    \begin{bmatrix}
    X_1 \\ 
    X_2 \\ 
    \zerom{a_i'}{n'-2\mm} \\
    \end{bmatrix}
    \in \mathbb{F}_{q'}^{m_i\times(n'-2\mm)}
    \bigg|
    X_1 \in \mathbb{F}_{q'}^{a_i\times(n'-2\mm)},
           X_2 \in \mathbb{F}_{q'}^{(m_i-a_i-a_i')\times(n'-2\mm)}
    }.
\end{align*}
For any $[M_1] \in C_1/C_2^{\perp}$ 
where $M_1\in\mathbb{F}_{q'}^{ (m_i-a_i-a_i')\times (n'-2\mm)}$,
define the quantum state $|[M_1]\rangle_b \in \cH_{\mathcal{C}}$ by
\begin{align*}
    \bitk*{[M_1]} :=&
    \frac{1}{\sqrt{|C_2^{\perp}|}} 
    \sum_{Y\in C_2^{\perp}}
    \bitk*{
    \begin{bmatrix}
    \zerom{a_i}{n'-2\mm}\\ 
    M_1 \\ 
    \zerom{a'_i}{n'-2\mm} 
    \end{bmatrix}
    +Y }  \nonumber
    =
    \begin{bmatrix}
    |\zerom{a_i}{n'-2\mm} \rangle_b \\ 
    |M_1\rangle_b \\ 
    |\zerom{a'_i}{n'-2\mm} \rangle_p \\
    \end{bmatrix}.
\end{align*}
With the above definitions, 
the code space is given as $\cH'_{\mathrm{code}} := \cH'_{\mathcal{C}2} = \cH'^{\otimes (m_i-a_i-a_i')\times (n'-2\mm)}$
and a pure state $|\phi\rangle 
\in \cH'_{\mathrm{code}}$
is encoded 
as a superposition of the states $|[M_1]\rangle_b$ in this CSS code by
\begin{align*}
\begin{bmatrix}
    |\zerom{a_i}{n'-2\mm} \rangle_b \\ 
    |\phi\rangle \\ 
    |\zerom{a_i'}{n'-2\mm} \rangle_p \\
\end{bmatrix}
    \in \cH_{\mathcal{C}}. 
\end{align*}

\section{Code Construction with Negligible Rate Shared Randomness} \label{sec:code}


In this section, we construct our code that allows a sender $S_i$ to transmit a state $\rho_i$ on $\cH'_{\mathrm{code}} = \cH'^{\otimes (m_i - a_i-a'_i)\times (n'-2m_i)}$ correctly to a receiver $T_i$ by $n$-use of the network
when the encoder and decoder share the negligible rate random variable $SR_i := (R_i,V_i)$.

The encoder and decoder are defined depending on the private randomness $U_{i,1}$ owned by encoder and the randomness $SR_i$ shared between the encoder and decoder.
These random variables are uniformly chosen from the values or matrices satisfying the following respective conditions:
the variable $R_i:=(R_{i,1}, R_{i,2}) \in\mathbb{F}_{q'}^{(m_i-a_i)\times m_i}\times\mathbb{F}_{q'}^{(m_i-a'_i)\times m_i}$ satisfies $\rank R_{i,1}= m_i-a_i$ and $\rank R_{i,2}= m_i-a'_i$,
the random variable $V_i := (V_{i,1},\ldots,V_{i,4m_i})$ consists of $4m_i$ values $V_{i,1},\ldots,V_{i,4m_i}\in\mathbb{F}_{q'}^{4m_i}$ 
and the random variable $U_{i,1}\in\mathbb{F}_{q'}^{m_i\times m_i}$ satisfies $\rank U_{i,1} = m_i$.

Next, we construct the encoder $\mathcal{E}_i^{SR_i,U_{i,1}}$ and decoder $\mathcal{D}_i^{SR_i}$.
Depending on $SR_i$ and $U_{i,1}$, the encoder $\mathcal{E}_i^{SR_i,U_{i,1}}$ of the sender $S_i$ is defined as an isometry channel from $\cH'_{\mathrm{code}}$ to $\cH_{S_i} = \cH'^{\otimes m_i\times n'}$.
Depending on $SR_i$, the decoder $\mathcal{D}_i^{SR_i}$ of the receiver $T_i$ is defined as a TP-CP map from $\cH_{T_i}=\cH'^{\otimes m_i\times n'}$ to $\cH'_{\mathrm{code}}$.
Note that the randomness $SR_i$ is shared between the encoder and the decoder.
Because $SR_i$ consists of $\alpha m_i(2m_i-a_i-a'_i+4)$ elements of $\mathbb{F}_{q}$,
the size of the shared randomness $SR_i $ 
is sublinear with respect to $n$ (i.e., negligible).

\subsection{Encoder $\mathcal{E}_i^{SR_i,U_{i,1}}$ of the sender $S_i$}
The encoder $\mathcal{E}_i^{SR_i,U_{i,1}}$ consists of three steps.
In the following, we describe the encoding of the state $\ket*{\phi}$ in $\cH'_{\mathrm{code}}$.

\begin{framed}
\noindent{\bf Step E1\quad}
The isometry map $U_{i,0}^{R_i}$ encodes the state $\ket*{\phi}$ with the CSS code defined in Subsection \ref{sec:css_code} and the quantum systems $\cH'_{\mathcal{A}}$ and $\cH'_{\mathcal{B}}$ as
\begin{align*}
\ket*{\phi_1} := U_{i,0}^{R_i} \ket*{\phi} =
\bitk*{\begin{bmatrix}
    \zerom{a_i}{m_i} \\ 
    \multicolumn{1}{c}{\multirow{2}{*}{$R_{i,1}$}} \\
    \multicolumn{1}{c}{} \end{bmatrix} }
\otimes
\phasek*{\begin{bmatrix} 
    \multicolumn{1}{c}{\multirow{2}{*}{$R_{i,2}$}} \\
    \multicolumn{1}{c}{} \\
    \zerom{a'_i}{m_i}  \end{bmatrix} }
\otimes
\begin{bmatrix}
\bitk*{\zerom{a_i}{m_i}}\\
\ket*{\phi}\\
\lvert\zerom{a_i'}{m_i}\rangle_p
\end{bmatrix}
\in \cH'_{\mathcal{A}}\otimes\cH'_{\mathcal{B}}\otimes \cH'_{\mathcal{C}} = \cH_{S_i}.
\end{align*}

\noindent{\bf Step E2\quad}
By quantum invertible linear operation $\leftop{U_{i,1}}$, the encoder maps $\ket*{\phi_1}$ to $\ket{\phi_2} := \leftop{U_{i,1}} \ket{\phi_1}$.

\noindent{\bf Step E3\quad}
From random variable $V_i = (V_{i,1},\ldots,V_{i,4m_i})$, 
define matrices $Q_{i,1;j,k}:=(V_{i,k})^j$, $Q_{i,2;j,k}:=(V_{i,m_i+k})^j$ for $1\leq j \leq n'-2m_i$, $1\leq k\leq m_i$,
and $Q_{i,3;j,k}:=(V_{i,2m_i+k})^j$, $Q_{i,4;j,k}:=(V_{i,3m_i+k})^j$ for $1\leq j,k \leq m_i$.
With these matrices, define the matrix $U_{i,2}^{\vsec}\in\mathbb{F}_{q'}^{n'\times n'}$ as
\begin{align*}
U_{i,2}^{\vsec}:=&
\!
\setlength\arraycolsep{1pt}
\left[
\!
\begin{array}{ccc}
I_{\mm} & \zerom{\mm}{\mm} & \zerom{\mm}{n'\!-\!2\mm} \\
Q_{i,3}^{\mathrm{T}}Q_{i,4} & I_{\mm} & \zerom{\mm}{n'\!-\!2\mm} \\
\zerom{n'\!-\!2\mm}{\mm} & \zerom{n'\!-\!2\mm}{\mm} & I_{n'\!-\!2\mm}
\end{array}
\!\!
\right]
 \!\!\cdot\!\!
\left[
\!
\begin{array}{ccc}
I_{\mm} & \zerom{\mm}{\mm} & \zerom{\mm}{n'\!-\!2\mm}  \\
\zerom{\mm}{\mm} & I_{\mm} & Q_{i,2}^{\mathrm{T}} \\
\zerom{n'\!-\!2\mm}{\mm} & \zerom{n'\!-\!2\mm}{\mm} & I_{n'\!-\!2\mm}
\end{array}
\!\!
\right]
\!\!\cdot\!\!
\left[\!
\begin{array}{ccc}
I_{\mm} & \zerom{\mm}{\mm} & \zerom{\mm}{n'\!-\!2\mm}  \\
\zerom{\mm}{\mm} & I_{\mm} & \zerom{\mm}{n'\!-\!2\mm} \\
Q_{i,1} & \zerom{n'\!-\!2\mm}{\mm} & I_{n'\!-\!2\mm}
\end{array}
\!\!\right],
\end{align*}
where $I_d$ is the identity matrix of size $d$.
By quantum invertible linear operation $\rightop{U_{i,2}^{V_i}}$, the encoder maps $\ket*{\phi_2}$ to
$\rightop{U_{i,2}^{V_i}} \ket*{\phi_2}$.
\end{framed}

By above three steps, the encoder $\mathcal{E}_i^{SR_i,U_{i,1}}$ is described as an isometry map
\begin{align*}
\mathcal{E}_i^{SR_i,U_{i,1}} : \ket*{\phi} \mapsto \rightop{U_{i,2}^{V_i}} \leftop{U_{i,1}}  U_{i,0}^{R_i} \ket*{\phi} \in \cH_{S_i}. 
\end{align*}

\subsection{Decoder $\mathcal{D}_i^{SR_i}$ of the receiver $T_i$} \label{subsec:dec}

Decoder $\mathcal{D}_i^{SR_i}$  consists of two steps.
In the following, we describe the decoding of the state $\ket*{\psi}\in\cH_{T_i}$. 

\begin{framed}
\noindent{\bf Step D1\quad}
Since $(U_{i,2}^{V_i})^{-1}$ can be constructed from shared randomness $V_i$ by
\begin{align*}
\!\!
(U_{i,2}^{\vsec})^{-1}
\!
=&
\!
\setlength\arraycolsep{1pt}
\!
\left[\!
\begin{array}{ccc}
I_{\mm} & \zerom{\mm}{\mm} & \zerom{\mm}{n'\!-\!2\mm} \\
\zerom{\mm}{\mm} & I_{\mm} & \zerom{\mm}{n'\!-\!2\mm} \\
-Q_{i,1} & \zerom{n'\!-\!2\mm}{\mm} & I_{n'\!-2\mm}
\end{array}
\!\!\right]
 \!\!\cdot\!\!
\left[\!
\begin{array}{ccc}
I_{\mm} & \zerom{\mm}{\mm} & \zerom{\mm}{n'\!-\!2\mm} \\
\zerom{\mm}{\mm}  & I_{\mm} & -Q_{i,2}^{\mathrm{T}} \\
\zerom{n'\!-\!2\mm}{\mm} & \zerom{n'\!-\!2\mm}{\mm} & I_{n'\!-\!2\mm}
\end{array}
\!\!\right]
 \!\!\cdot\!\!
\left[\!
\begin{array}{ccc}
I_{\mm} & \zerom{\mm}{\mm} & \zerom{\mm}{n'-2\mm} \\
-Q_{i,3}^{\mathrm{T}}Q_{i,4} &  I_{\mm} & \zerom{\mm}{n'\!-\!2\mm} \\
\zerom{n'\!-\!2\mm}{\mm} & \zerom{n'\!-\!2\mm}{\mm} & I_{n'\!-\!2\mm}
\end{array}
\!\!\right],
\end{align*}
the decoder applies the reverse operation $\rightop{U_{i,2}^{V_i}}^{\dagger}=\rightop{(U_{i,2}^{V_i})^{-1}}$ of Step E3 as $\ket*{\psi_1} := \rightop{U_{i,2}^{V_i}}^{\dagger}\ket*{\psi}$.

\noindent{\bf Step D2\quad}
Perform the bit and phase basis measurements on $\cH'_{\mathcal{A}}$ and $\cH'_{\mathcal{B}}$, respectively, 
and let $O_{i,1}, O_{i,2}\in \mathbb{F}_{q'}^{m_i\times m_i}$ be the respective measurement outcomes.
By Gaussian elimination, find invertible matrices $D_{i,1}^{R_{\!i,\!1},O_{\!i,\!1}}, D_{i,2}^{R_{\!i,\!2},O_{\!i,\!2}} \in \mathbb{F}_{q'}^{m_i\times m_i}$ satisfying
\begin{align}
P_{\mathcal{W}_{i,1}} D_{i,1}^{R_{\!i,\!1},O_{\!i,\!1}} O_{i,1} &= 
\begin{bmatrix}
    \zerom{a_i}{m_i} \\ 
    \multicolumn{1}{c}{\multirow{2}{*}{$R_{i,1}$}} \\
    \multicolumn{1}{c}{} \end{bmatrix}, \quad
P_{\mathcal{W}_{i,2}} D_{i,2}^{R_{\!i,\!2},O_{\!i,\!2}} O_{i,2} = 
\begin{bmatrix} 
    \multicolumn{1}{c}{\multirow{2}{*}{$R_{i,2}$}} \\
    \multicolumn{1}{c}{} \\
    \zerom{a'_i}{m_i}  \end{bmatrix}. \label{eq:equation}
\end{align}
where $P_\mathcal{W}$ is the projection from $\mathbb{F}_{q'}^{m_i}$ to the subspace $\mathcal{W}$,
the subspace $\mathcal{W}_{i,1}$ consists of the vectors whose $1$-st, \ldots , $a_i$-th elements are zero
and the subspace $\mathcal{W}_{i,2}$ consists of the vectors whose $(m_i-a'_i+1)$-st, \ldots , $m_i$-th elements are zero.
The case of non-existence of $D_{i,1}^{R_{\!i,\!1},O_{\!i,\!1}}$ nor $D_{i,2}^{R_{\!i,\!2},O_{\!i,\!2}}$ 
means decoding failure, which implies that the decoder performs no more operations.
Also, 
when $D_{i,1}^{R_{\!i,\!1},O_{\!i,\!1}}$ and $D_{i,2}^{R_{\!i,\!2},O_{\!i,\!2}}$ are not determined uniquely,
the decoder chooses $D_{i,1}^{R_{\!i,\!1},O_{\!i,\!1}}$ and $D_{i,2}^{R_{\!i,\!2},O_{\!i,\!2}}$ deterministically depending on $O_{i,1},R_{i,1}$ and $O_{i,2}, R_{i,2}$, respectively.

Based on $D_{i,1}^{R_{\!i,\!1},O_{\!i,\!1}}$ and $D_{i,2}^{R_{\!i,\!2},O_{\!i,\!2}}$ found by \eqref{eq:equation},
the decoder applies $\leftop{D_{i,1}^{R_{\!i,\!1},O_{\!i,\!1}}}$ and $\leftop{\ph{D_{i,2}^{R_{\!i,\!2},O_{\!i,\!2}}}}$ consecutively to $\ket*{\psi_1}$,
and the resultant state on $\cH_{\mathrm{code}}$ is the output of Step D2.
Then, Step D2 is written as the following TP-CP map $D_i^{R_i}$:
    \begin{align*}
    D_i^{R_i}(\kb{\psi_1}) :=&
    \Tr_{\mathcal{C}1,\mathcal{C}3}
    \!\!\!\!
    \sum_{O_{\!i,\!1}\!,O_{\!i,\!2}\in\mathbb{F}_{q'}^{m_i\times \mm}}
    \!\!\!\!
    U_D^{R_i\!,O_{i,\!1}\!,O_{i,\!2}}
    \sigma_{O_{\!i,\!1}\!,O_{\!i,\!2}}
    (U_D^{R_i\!,O_{i,\!1}\!,O_{i,\!2}})^{\dagger},
    \end{align*}
    where the matrices $U_D^{R_i\!,O_{i,\!1}\!,O_{i,\!2}}$ and $\sigma_{O_{\!i,\!1}\!,O_{\!i,\!2}}$ are defined as
    \begin{align*}
    U_D^{R_i\!,O_{i,\!1}\!,O_{i,\!2}}:=&\leftop{\ph{D_{i,2}^{R_{\!i,\!2}\!,O_{\!i,\!2}}}} \leftop{D_{i,1}^{R_{\!i,\!1}\!,O_{\!i,\!1}}},
    \quad
    \sigma_{O_{\!i,\!1}\!,O_{\!i,\!2}} :=
    \Tr_{\mathcal{A},\mathcal{B}} 
    \kb{\psi_1}
    \paren*{|O_{\!i,\!1}\rangle_{b}{}_b\langle O_{\!i,\!1}|
    \otimes
    |O_{\!i,\!2}\rangle_{p}{}_p\langle O_{\!i,\!2}|
    \otimes I_{\mathcal{C}}},
    \end{align*}
    with the identity operator $I_{\mathcal{C}}$ on $\cH_{\mathcal{C}}$.
\end{framed}
By above two steps, the decoder $\mathcal{D}_i^{SR_i}$ is described as 
\begin{align*}
\mathcal{D}_i^{SR_i}(\kb{\psi}) := D_i^{R_{i}} \paren*{\rightop{U_{i,2}^{V_i}}^{\dagger} \kb{\psi} \rightop{U_{i,2}^{V_i}} }.
\end{align*}
Since the size of the shared randomness $SR_i$ is sublinear with respect to $n$, our code is implemented with negligible rate shared randomness.

\section{Correctness of Our Code}  \label{sec:analysis}

In this section, we confirm that our code correctly transmits the state from the sender $S_i$ to the receiver $T_i$.
As is mentioned in Section \ref{sec:main_result}, we show the condition $n(1-F_e^2(\rho_{mix},\kappa_i))\to 0$ which implies the correctness of our code.

First, we describe the quantum code protocol $\kappa_i$ from $S_i$ to $T_i$, 
which is an integration of the encoding, transmission, and decoding.
The encoding and decoding in $\kappa_i$ is given by the probabilistic mixture of the code in Section \ref{sec:code} depending on the uniformly chosen random variables $SR_i$ and $U_{i,1}$.
Then, the code protocol $\kappa_i$ is written as, for the state $\rho_i$ on $\cH'_{\mathrm{code}}$,
\begin{align*}
\kappa_i(\rho_i) :=&  \sum_{SR_i,U_{i,1}} \frac{1}{N}\mathcal{D}_i^{SR_i} \paren*{\Tr_{T_1,\ldots,T_{i-1},T_{i+1},\ldots,T_{\p}} \lefto{K} \paren*{\mathcal{E}_i^{SR_i,U_{i,1}}\paren*{\rho_i}\otimes \rho_{i^c}} \lefto{K}^\dagger}, 
\end{align*}
where
$\rho_{i^c}$ is the state in $\cH_{S_1}\otimes \cdots \otimes \cH_{S_{i-1}}\otimes \cH_{S_{i+1}}\otimes\cdots \otimes \cH_{S_\p}$ of senders other than $S_i$,
and 
$N :=        {q'}^{4m_i}
        + |\{U_{i,1}\in\mathbb{F}_{q'}^{m_i\times m_i} | \rank U_{i,1}=m_i \}|
        + |\{R_{i,1}\in\mathbb{F}_{q'}^{(m_i-a_i)\times m_i} | \rank R_{i,1} = m_i-a_i \}|
        + |\{R_{i,2}\in\mathbb{F}_{q'}^{(m_i-a_i')\times m_i} | \rank R_{i,1} = m_i-a_i' \}|$.


As explained in \cite[Section IV]{SM18}, 
$1-F_e^2(\rho_{mix}, \kappa_i)$ is upper bounded by the sum of the bit error probability and the phase error probability.
The bit error probability is the probability that a bit basis state $\bitk*{X}\in\cH'_{\mathrm{code}}$ is sent but the bit basis measurement outcome on the decoder output is not $X$.
In the similar way, the phase error probability is defined for the phase basis.
%
%
We will show in Subsections \ref{subsec:bit} and \ref{subsec:phase} that 
the bit and phase error probabilities are upper bounded by $O\paren*{\max \Big\{\frac{1}{q'} , \frac{(n')^{m_i}}{(q')^{m_i-a_i}} \Big\}}$ and $O\paren*{\max \Big\{\frac{1}{q'} , \frac{(n')^{m_i}}{(q')^{m_i-a'_i}} \Big\}}$, respectively.
Therefore, we have
\begin{align}
n(1-F_e^2(\rho_{mix}, \kappa_i)) \leq nO\paren*{\max \Big\{\frac{1}{q'} , \frac{(n')^{m_i}}{(q')^{m_i-\max\{a_i,a_i'\}}} \Big\}}. \label{eq:upperbounds}
\end{align}
Since $q'$ is taken in Section \ref{sec:prelim} to satisfy $\frac{n\cdot (n')^{m_i}}{(q')^{m_i-\max\{a_i,a_i'\}}}\to 0$, 
the RHS of \eqref{eq:upperbounds} converges to $0$ and therefore $n(1-F_e^2(\rho_{mix}, \kappa_i))\to 0$.
This completes the proof of Theorem \ref{theo:main1}.

\subsection{Notation and Lemmas for Bit and Phase Error Probabilities} \label{sec:bit_phase_lemmas}
In this subsection, we prepare a notation and lemmas for proving the upper bounds of the bit and phase error probabilities.
The upper bounds of these probabilities are shown separately in Subsections \ref{subsec:bit} and \ref{subsec:phase}.  

We introduce the notation 
$X := (X^{\mathcal{A}}, X^{\mathcal{B}}, X^{\mathcal{C}}) \in \mathbb{F}_{q'}^{k\times m_i}\times \mathbb{F}_{q'}^{k\times m_i}\times \mathbb{F}_{q'}^{k\times (n'-2m_i)}$
for $X\in\mathbb{F}_{q'}^{k\times n'}$ with arbitrary positive integer $k$.
Also, we prepare the following lemmas.
\begin{lemm} \label{lemm:prob_bit}
For integers $d_0 \geq d_1+d_2$, 
let $\mathcal{W}_1\subset \mathbb{F}_{q'}^{d_0}$ be a $d_1$-dimensional subspace and
$\mathcal{W}_2\subset \mathbb{F}_{q'}^{d_0}$ be a $d_2$-dimensional subspace.
Assume the following three conditions.
\begin{description}
\item[${(\Gamma 1)}$] $\mathcal{W}_1\cap\mathcal{W}_2 = \{\zerom{d_0}{1}\}$.
\item[${(\Gamma 2)}$] Let $\xm\geq d_1+d_2$. The vectors $x_1,\ldots, x_{\xm} \in \mathcal{W}_1$ and $y_1,\ldots, y_{\xm} \in \mathcal{W}_2$ satisfy
\begin{align*}
\myspan\paren*{ (x_1,y_1),\ldots, (x_{\xm},y_{\xm}) } = \mathcal{W}_1\oplus \mathcal{W}_2.
\end{align*}
\item[${(\Gamma 3)}$] Let $W_1'\subset \mathbb{F}_{q'}^{d_0}$ be a $d_1$-dimensional subspace and $r_1,\ldots, r_{\xm}\in \mathcal{W}_1'$.
There exists an invertible linear map $A: \mathcal{W}_1' \to \mathcal{W}_1 $ which maps 
\begin{align*}
 [x_1,\ldots, x_{\xm}] = A [r_1,\ldots, r_{\xm}].
\end{align*}
\end{description}

Then, the following two statements hold.
\begin{description}
\item[${(\Delta 1)}$] There exists invertible linear map $D:\mathbb{F}_{q'}^{d_0} \to \mathbb{F}_{q'}^{d_0}$ that
\begin{align}
P_{\mathcal{W}_1'} D[ (x_1,y_1),\ldots, (x_{\xm},y_{\xm}) ]  = A^{-1} [x_1,\ldots, x_{\xm}] = [r_1,\ldots, r_{\xm}]. \label{eq:prob_bit_i}
\end{align}
\item[${(\Delta 2)}$] For the above linear map $D$, it holds for any $x\in\mathcal{W}_1$ and $y\in\mathcal{W}_2$ that
\begin{align}
P_{\mathcal{W}_1'} D (x,y) =  A^{-1}x.  \label{eq:prob_bit_b}
\end{align}
\end{description}
\begin{IEEEproof}
First, we show the item ${(\Delta 1)}$.
Let $\mathcal{W}_3$ be a subspace of $\mathbb{F}_{q'}^{d_0}$ that satisfies $\mathcal{W}_1\oplus\mathcal{W}_2\oplus\mathcal{W}_3 = \mathbb{F}_{q'}^{d_0}$.
If $D$ is defined as 
$D|_{\mathcal{W}_1} = A^{-1}$ and
$D|_{\mathcal{W}_2\oplus\mathcal{W}_3} (\mathcal{W}_2\oplus\mathcal{W}_3)= \mathcal{W}_1'^{\perp}$,
we obtain \eqref{eq:prob_bit_i}, i.e., ${(\Delta 1)}$ from
\begin{align*}
P_{\mathcal{W}_1'}D((x_i,y_i)) = P_{\mathcal{W}_1'}(D|_{\mathcal{W}_1} (x_i)+ D|_{\mathcal{W}_2\oplus\mathcal{W}_3}(y_i)) = A^{-1} x_i = r_i.
\end{align*}
Next, we show the item ${(\Delta 2)}$.
Since arbitrary $(x,y)\in\mathcal{W}_1\oplus\mathcal{W}_2$ is spanned by $(x_1,y_1),\ldots, (x_{\xm},y_{\xm})$,
Eq. \eqref{eq:prob_bit_i} implies \eqref{eq:prob_bit_b}, which yields ${(\Delta 2)}$.
\end{IEEEproof}
\end{lemm}

\begin{lemm}[{\cite[Lemma 7.1]{SM18}}] \label{lemm:space}
For integers $d_a\geq d_b+d_c$,
fix a $d_b$-dimensional subspace $\mathcal{W}\subset \mathbb{F}_{q'}^{d_a}$,
and randomly choose a $d_c$-dimensional subspace $\mathcal{R}\subset \mathbb{F}_{q'}^{d_a}$ with the uniform distribution.
Then, we have
\begin{align*}
\Pr[\mathcal{W}\cap \mathcal{R}=\{\zerom{d_a}{1}\}] = 1-O({q'}^{d_b+d_c-d_a-1}).
\end{align*}
\end{lemm}

\begin{lemm} \label{lemm:space2}
For $d \geq d'$,
\begin{align*}
\Pr\bparen*{ \rank [t_1,\ldots, t_d ] = d' \mathrel{\Big|} t_1,\ldots, t_d \in \mathbb{F}_{q'}^{d'} } \geq 1-O\paren*{\frac{1}{q'}}.
\end{align*}
\begin{IEEEproof}
From $d \geq d'$, we have
\begin{align}
 \Pr\bparen*{ \rank[t_1,\ldots, t_d ]=d' \mathrel{\Big|} t_1,\ldots, t_d \in \mathbb{F}_{q'}^{d'} } 
\geq \Pr\bparen*{ \rank[t_1,\ldots, t_{d'}]=d' \mathrel{\Big|} t_1,\ldots, t_{d'} \in \mathbb{F}_{q'}^{d'} }. \label{eq:prob_proof}
\end{align}
On the other hand, 
the RHS of \eqref{eq:prob_proof} is equivalent to the probability to choose $d'$ independent vectors in $\mathbb{F}_{q'}^{d'}$:
\begin{align*}
     &\Pr\bparen*{ \rank[t_1,\ldots, t_{d'}]=d' \mathrel{\Big|}  t_1,\ldots, t_{d'} \in \mathbb{F}_{q'}^{d'} }
    =\frac{{q'}^{d'}}{{q'}^{d'}}\cdot\frac{{q'}^{d'}-q'}{{q'}^{d'}}
     \cdots 
     \frac{{q'}^{d'}-{q'}^{d'-1}}{{q'}^{d'}}
    =1 - O\paren*{\frac{1}{q'}}.  
\end{align*}
By combining the above inequality and equality, we have the lemma. 
\end{IEEEproof}
\end{lemm}

\begin{lemm}[{\cite[Lemmas 7.2 and 7.4]{SM18}}]    \label{lemm:max_zero_prob}
For the random matrix $U_{i,2}^{V_i}$ defined in Step E3, we have
\begin{align*}
 &\max_{\zerom{n'}{1}\neq x\in\mathbb{F}_{q'}^{n'}} \!\!\!\!\!\!\Pr[ x^{\mathrm{T}} ((U_{i,2}^{V_i})^{-1})^{\mathcal{A}} \! = \! \zerom{1}{\mm} ] \leq \Big(\frac{n'\!-\!2\mm}{q'}\Big)^{\!\mm},
 \\
 &\max_{\zerom{n'}{1}\neq x\in\mathbb{F}_{q'}^{n'}}\!\!\!\!\!\! \Pr[ x^{\mathrm{T}} ((\ph{U_{i,2}^{V_i}})^{-1})^{\mathcal{B}} \!=\! \zerom{1}{\mm} ] \leq \Big(\frac{n'\!-\!2\mm}{q'}\Big)^{\!\mm}. 
\end{align*}
\end{lemm}

\subsection{Bit Error Probability} \label{subsec:bit}
In this subsection, we show that 
when arbitrary bit basis state $\bitk{M}\in\cH'_{\mathrm{code}}$ is the input state of the sender $S_i$, 
the original message $M$ is correctly recovered with probability at least $1-O\paren*{\max \Big\{\frac{1}{q'} , \frac{(n')^{m_i}}{(q')^{m_i-a_i}} \Big\}}$.

\noindent{\it Step 1:\quad}
We derive a necessary condition for correct decoding of the original message $M$ in bit basis. 
When arbitrary bit basis state $\bitk{M}\in\cH'_{\mathrm{code}}$ is the input state of the sender $S_i$, the encoded state is written as 
\begin{align}
\mathcal{E}_i^{SR_i,R_i}(\bitk{M}) =
\sum_{\bar{E}_1 \in \mathbb{F}_{q'}^{m_i\times m_i},
\bar{E}_2 \in \mathbb{F}_{q'}^{a'_i\times (n'-2m_i)}}
\bitk*{ U_{i,1} 
    \left[\begin{array}{ccc}
    \zerom{a_i}{\mm} &  \multicolumn{1}{c}{\multirow{3}{*}{$\bar{E}_1$}}  & \zerom{a_i}{n'-2\mm}\\
    \multicolumn{1}{c}{\multirow{2}{*}{$R_{i,1}$}} & \multicolumn{1}{c}{} & M \\
    \multicolumn{1}{c}{} & \multicolumn{1}{c}{}  & \bar{E}_2\\
    \end{array}\right]
    U_{i,2}^{V_i}   }, \nonumber
\end{align}
where we ignore the normalizing factors and phase factors.

Note that bit state measurement on network output system $\cH_{T_i} = \cH'^{\otimes m_i \times n'_i}$ commutes with the decoding operation $\mathcal{D}_i^{SR_i}$ on $\cH_{T_i}$.
Therefore, in the analysis of the bit error probability, we take the method to perform bit state measurement to $\cH_{T_i}$ first, and then apply the decoding operation corresponding to $\mathcal{D}_i^{SR_i}$,
instead of decoding first and performing bit state measurement. 

By performing the bit basis measurement to the network output $\sigma_{T_i} = \kappa_i(|M\rangle_{bb}\langle M|)$, 
we have the following measurement outcome $Y$:
\begin{align}
Y = K_{i,i} U_{i,1} 
    \left[\begin{array}{ccc}
    \zerom{a_i}{\mm} &  \multicolumn{1}{c}{\multirow{3}{*}{$\bar{E}_1$}}  & \zerom{a_i}{n'-2\mm}\\
    \multicolumn{1}{c}{\multirow{2}{*}{$R_{i,1}$}} & \multicolumn{1}{c}{} & M \\
    \multicolumn{1}{c}{} & \multicolumn{1}{c}{}  & \bar{E}_2\\
    \end{array}\right]
    U_{i,2}^{V_i} 
  + K_{i^c} Z,  \nonumber
\end{align}
where 
$\bar{E}_1 \in \mathbb{F}_{q'}^{m_i\times m_i}$,
$\bar{E}_2 \in \mathbb{F}_{q'}^{a'_i\times (n'-2m_i)}$
and 
$Z \in \mathbb{F}_{q'}^{(m-m_i)\times n'}$.
By Step D1, $Y$ is decoded to 
\begin{align}
\bar{Y} =
Y(U_{i,2}^{V_i})^{-1} = 
K_{i,i} U_{i,1} 
    \begin{bmatrix}
    \zerom{a_i}{\mm} &  \multicolumn{1}{c}{\multirow{3}{*}{$\bar{E}_1$}}  & \zerom{a_i}{n'-2\mm}\\
    \multicolumn{1}{c}{\multirow{2}{*}{$R_{i,1}$}} & \multicolumn{1}{c}{} & M \\
    \multicolumn{1}{c}{} & \multicolumn{1}{c}{}  & \bar{E}_2\\
    \end{bmatrix}
    + K_{i^c} Z (U_{i,2}^{V_i})^{-1} . \nonumber
\end{align}
The measurement outcome $O_{i,1}$ in Step D2 is
\begin{align*}
O_{i,1} = \bar{Y}^{\mathcal{A}} =
K_{i,i} U_{i,1} 
    \begin{bmatrix}
    \zerom{a_i}{\mm} \\
    \multicolumn{1}{c}{\multirow{2}{*}{$R_{i,1}$}} \\
    \multicolumn{1}{c}{} 
    \end{bmatrix}
    +
    (K_{i^c} Z (U_{i,2}^{V_i})^{-1})^{\mathcal{A}}.
\end{align*}
Since the decoder knows $O_{i,1}$ and $R_{i,1}$, the matrix $D_{i,1}^{R_{\!i,\!1},O_{\!i,\!1}}$ is found by Gaussian elimination to the left equation of \eqref{eq:equation} which is written as
\begin{align}
P_{\mathcal{W}_{i,1}} D_{i,1}^{R_{\!i\!,1},O_{\!i\!,1}} O_{i,1} &=
P_{\mathcal{W}_{i,1}} D_{i,1}^{R_{\!i\!,1},O_{\!i\!,1}} 
    \paren*{
    K_{i,i} U_{i,1} 
    \begin{bmatrix}
    \zerom{a_i}{\mm} \\
    \multicolumn{1}{c}{\multirow{2}{*}{$R_{i,1}$}} \\
    \multicolumn{1}{c}{} 
    \end{bmatrix}
    +
    (K_{i^c} Z (U_{i,2}^{V_i})^{-1})^{\mathcal{A}}}
=
\begin{bmatrix}
    \zerom{a_i}{m_i} \\ 
    \multicolumn{1}{c}{\multirow{2}{*}{$R_{i,1}$}} \\
    \multicolumn{1}{c}{} \end{bmatrix}. \label{eq:Gaussian_elim}
\end{align}
Therefore, 
if the matrix $D_{i,1}^{R_{\!i,\!1},O_{\!i,\!1}}$ derived in \eqref{eq:Gaussian_elim}
satisfies the following equation
\begin{align}
P_{\mathcal{W}_{\!i\!,1}} \!D_{i\!,1}^{R_{i\!,1},O_{\!i\!,1}} \bar{Y}^{\mathcal{C}} =
P_{\mathcal{W}_{\!i\!,1}} \!D_{i\!,1}^{R_{i\!,1},O_{\!i\!,1}}
    \paren*{
    K_{i\!,i} U_{i\!,1} \!
    \begin{bmatrix}
    \zerom{a_i}{n'\!-\!2\mm}\\
    M \\
    \bar{E}_2
    \end{bmatrix}
    +
    (K_{i^c} Z (U_{i\!,2}^{\!V_i})^{-1})^{\mathcal{C}}}
 =
    \begin{bmatrix}
    \zerom{a_i}{n'\!-\!2\mm}\\
    M \\
    \bar{E}_2
    \end{bmatrix}, \label{eq:prob_bit}
\end{align}
the original message $M$ is correctly recovered. 

\noindent{\it Step 2:\quad}
In the next step, we show that
the conditions ${(\Gamma 1)}$, ${(\Gamma 2)}$ and ${(\Gamma 3)}$ of
Lemma \ref{lemm:prob_bit} in the following case imply Eq. \eqref{eq:prob_bit}; 
\begin{gather*}
\mathcal{W}_1:=\col\paren*{K_{i,i} U_{i,1}
    \begin{bmatrix}
    \zerom{a_i}{\mm} \\
    \multicolumn{1}{c}{\multirow{2}{*}{$R_{i,1}$}} \\
    \multicolumn{1}{c}{} 
    \end{bmatrix}}, \quad
\mathcal{W}_2:=\col\paren*{ K_{i^c} Z (U_{i,2}^{V_i})^{-1}}, 
\quad \mathcal{W}_1':= \mathcal{W}_{i,1},
\quad \xm:=m_i, \\
[x_1,\ldots, x_{\xm}] := K_{i,i} U_{i,1}
    \begin{bmatrix}
    \zerom{a_i}{\mm} \\
    \multicolumn{1}{c}{\multirow{2}{*}{$R_{i,1}$}} \\
    \multicolumn{1}{c}{} 
    \end{bmatrix}, \quad
[y_1,\ldots, y_{\xm}] := (K_{i^c} Z (U_{i,2}^{V_i})^{-1})^{\mathcal{A}}, \\
[r_1,\ldots, r_{\xm}] := 
    \begin{bmatrix}
    \zerom{a_i}{\mm} \\
    \multicolumn{1}{c}{\multirow{2}{*}{$R_{i,1}$}} \\
    \multicolumn{1}{c}{}
    \end{bmatrix},\quad
     A := (K_{i,i} U_{i,1})|_{\mathcal{W}_1'},\quad
(d_0,d_1,d_2) := (m_i, m_i-a_i, \rank K_{i^c} Z),
\end{gather*}
where $\col(T)$ of the matrix $T$ is the column space of $T$
and $\mathcal{W}_{i,1}$ is defined in Step D2 of Subsection \ref{subsec:dec}.

Applying Lemma \ref{lemm:prob_bit}, we show that Eq. \eqref{eq:prob_bit} holds if the conditions ${(\Gamma 1)}$, ${(\Gamma 2)}$ and ${(\Gamma 3)}$ are satisfied.
Assume that ${(\Gamma 1)}$, ${(\Gamma 2)}$ and ${(\Gamma 3)}$ are satisfied. 
Then, the condition ${(\Delta 1)}$ holds which implies the existence of $D_{i,1}^{R_{\!i,\!1},O_{\!i,\!1}}$ in \eqref{eq:Gaussian_elim}.
Moreover, ${(\Delta 2)}$ implies that
for any $r\in\mathcal{W}_1', y\in \mathcal{W}_2$ and $x = K_{i,i} U_{i,1}r \in\mathcal{W}_1$, it holds
\begin{align*}
 P_{\mathcal{W}_1'}D_{i,1}^{R_{\!i,\!1},O_{\!i,\!1}}(x+y) = A^{-1}x = \paren*{(K_{i,i} U_{i,1})|_{\mathcal{W}_1'}}^{-1} (K_{i,i} U_{i,1}r)  = r,
\end{align*}
and this yields \eqref{eq:prob_bit}.

\noindent{\it Step 3:\quad}
In the third step, 
we show that the relations ${(\Gamma 1)}$, ${(\Gamma 2)}$ and ${(\Gamma 3)}$ hold
at least with probability $1-O\Big(\max \Big\{\frac{1}{q'} , \frac{(n')^{m_i}}{(q')^{m_i-a_i}} \Big\}\Big)$, which 
proves the desired statement by combining the conclusion of Steps 1 and 2.

\noindent{\it Step 3-1:\quad}
In this substep, we show that the probability satisfying ${(\Gamma 1)}$, ${(\Gamma 2)}$ and ${(\Gamma 3)}$ is obtained by
\begin{align}
 \Pr[{(\Gamma 1)}\cap{(\Gamma 2)}\cap{(\Gamma 3)}] 
                =& \Pr[{(\Gamma 1)}] \cdot \Pr[{(\Gamma 2')}] \cdot \Pr[{(\Gamma 2)}|{(\Gamma 2')}\cap{(\Gamma 1)}], \label{eq:prob_process}
\end{align}
where the condition ${(\Gamma 2')}$ is given as
\begin{description}
\item[${(\Gamma 2')}$] $ \rank K_{i^c} Z ((U_{i,2}^{V_i})^{-1})^{\mathcal{A}} =\rank K_{i^c} Z $.
\end{description}
Eq. \eqref{eq:prob_process} is derived by the following reductions:
\begin{align*}
 &\Pr[{(\Gamma 1)}\cap{(\Gamma 2)}\cap{(\Gamma 3)}] 
                \stackrel{(a)}{=} \Pr[{(\Gamma 1)}\cap{(\Gamma 2)}] 
                \stackrel{(b)}{=}  \Pr[{(\Gamma 1)}] \cdot \Pr[{(\Gamma 2)}|{(\Gamma 1)}] \\
                \stackrel{(c)}{=}&  \Pr[{(\Gamma 1)}] \cdot \Pr[{(\Gamma 2)}\cap{(\Gamma 2')}|{(\Gamma 1)}] 
                 \stackrel{(d)}{=} \Pr[{(\Gamma 1)}] \cdot \Pr[{(\Gamma 2')}|{(\Gamma 1)}] \cdot \Pr[{(\Gamma 2)}|{(\Gamma 2')}\cap{(\Gamma 1)}]\\
                \stackrel{(e)}{=}& \Pr[{(\Gamma 1)}] \cdot \Pr[{(\Gamma 2')}] \cdot \Pr[{(\Gamma 2)}|{(\Gamma 2')}\cap{(\Gamma 1)}].
\end{align*}
The equality $(a)$ follows from the fact that ${(\Gamma 3)}$ is always satisfied for $A$ defined in Step 2,
and 
$(b)$ and $(d)$ are trivial.
$(c)$ is obtained because ${(\Gamma 2')}$ is a necessary condition for ${(\Gamma 2)}$.
Since $\myspan(y_1,\ldots,y_{\xm})=\mathcal{W}_2$ is a necessary condition for ${(\Gamma 2)}$ in Lemma \ref{lemm:prob_bit},
the condition ${(\Gamma 2')}$  is also necessary for ${(\Gamma 2)}$ from
\begin{align*}
\rank K_{i^c} Z ((U_{i,2}^{V_i})^{-1})^{\mathcal{A}}
&\!=\! \rank (K_{i^c} Z (U_{i,2}^{V_i})^{-1})^{\mathcal{A}}
\!=\! \dim\! \myspan(y_1,\ldots,y_{\xm})\\
&\!=\!\dim \mathcal{W}_2 
\!=\! \rank K_{i^c} Z (U_{i,2}^{V_i})^{-1} 
\!=\! \rank K_{i^c} Z .
\end{align*}
The equality $(e)$ follows from the fact that ${(\Gamma 1)}$ and ${(\Gamma 2')}$ are independent, which will be shown by $\Pr[{(\Gamma 1)}|{(\Gamma 2')}] = \Pr[{(\Gamma 1)}]$ in Step 3-2.

\noindent{\it Step 3-2:\quad}
In this step, we prove $\Pr[{(\Gamma 1)}]\geq 1-O\paren*{1/{q'}}$ and $\Pr[{(\Gamma 1)}|{(\Gamma 2')}] = \Pr[{(\Gamma 1)}]$.
Fix $R_{i,1}$ and $U_{i,2}^{V_i}$.
Then, $\mathcal{W}_1$ is randomly chosen $d_1$-dimensional subspace under uniform distribution
and $\mathcal{W}_2$ is fixed $d_2$-dimensional subspace.
Therefore, Lemma \ref{lemm:space} can be applied with $(d_a,d_b,d_c,\mathcal{W}):= (d_0,d_2, d_1,\mathcal{W}_2)$ and 
$\Pr[{(\Gamma 1)}] = 1-O\paren{ {q'}^{d_2 + d_1 - d_0 -1 } } \geq 1-O\paren*{1/{q'}}$.
On the other hand, since $\Pr[{(\Gamma 1)}]$ does not depend on $U_{i,2}^{V_i}$ but $\Pr[{(\Gamma 2)}]$ depends only on $U_{i,2}^{V_i}$, we have $\Pr[{(\Gamma 1)}|{(\Gamma 2')}] = \Pr[{(\Gamma 1)}]$. 

\noindent{\it Step 3-3:\quad}
In this step, we show $\Pr[{(\Gamma 2')}] \geq 1-\frac{n'^{m_i}}{{q'}^{m_i-a_i} }$.
The condition ${(\Gamma 2')}$ holds if and only if $x^{\mathrm{T}} K_{i^c} Z ((U_{i,2}^{V_i})^{-1})^{\mathcal{A}} \neq \zerom{1}{m_i}$ for any vector $x\in\mathbb{F}_{q'}^{m_i}$ satisfying $x^{\mathrm{T}} K_{i^c} Z \neq \zerom{1}{n'}$ (considering $K_{i^c}$, $Z$ and $((U_{i,2}^{V_i})^{-1})^{\mathcal{A}}$ as linear maps on row vector spaces, this is equivalent that $((U_{i,2}^{V_i})^{-1})^{\mathcal{A}}$ has trivial kernel $\{\zerom{1}{n'}\}$ for the image of $K_{i^c} Z $).
Therefore, by applying Lemma \ref{lemm:max_zero_prob} for all distinct $x^{\mathrm{T}} K_{i^c} Z$ which is not zero vector, we have
\begin{align*}
\Pr[{(\Gamma 2')}] \geq 1- {q'}^{\rank K_{i^c} Z} \paren*{\frac{n'-2m_i}{q'}}^{m_i}\geq 1-{q'}^{a_i}\paren*{\frac{n'-2m_i}{q'}}^{m_i} \geq 1-\frac{n'^{m_i}}{{q'}^{m_i-a_i} }.
\end{align*}

\noindent{\it Step 3-4:\quad}
Now we evaluate the probability $\Pr[{(\Gamma 2)}|{(\Gamma 2')}\cap{(\Gamma 1)}]\geq 1-O(1/{q'}^{-1})$.
Fix the random variable $U_{i,2}^{V_i}$ so that ${(\Gamma 2')}$ holds in the following.
Define matrices 
$T_x = [x_{i(1)},\ldots,x_{i(d_1+d_2)}] $,
$T_y = [y_{i(1)},\ldots,y_{i(d_1+d_2)}] $ and 
$T = T_x + T_y \in\mathbb{F}_{q'}^{d_0\times (d_1+d_2)}$ where $i:\{1,\ldots,d_1+d_2\}\to \{1,\ldots,\xm\}$ is an injective index function such that $y_{i(1)},\ldots, y_{i(d_2)}$ are linearly independent i.e., $\rank T_y = d_2$.
Then, we have
\begin{align*}
\Pr&\bparen*{{(\Gamma 2)}|{(\Gamma 2')}\!\cap\!{(\Gamma 1)} } 
                          \!\geq\!  \Pr[ \myspan\paren*{ (x_{i(1)},\!y_{i(1)}),\!\ldots, (x_{i(d_1+d_2)},y_{i(d_1+d_2)}) } \!=\! \mathcal{W}_1\!\oplus\! \mathcal{W}_2 \!\mid\!{(\Gamma 2')}\!\cap\!{(\Gamma 1)}] \nonumber\\
                          \stackrel{(a)}{=}& \Pr\bparen*{\rank T = d_1\!+\!d_2\mid{(\Gamma 2')}\!\cap\!{(\Gamma 1)}} 
                                                    = \Pr\bparen*{\ker T = \{\zerom{d_1\!+\!d_2}{1}\}\mid{(\Gamma 2')}\cap{(\Gamma 1)}} \nonumber \\
                          \stackrel{(b)}{=}& \Pr\bparen*{\ker T_x \cap \ker T_y = \{\zerom{d_1+d_2}{1}\}\mid {(\Gamma 2')}\cap{(\Gamma 1)}},
\end{align*}
where $(a)$ follows from $\myspan\paren*{ (x_{i(1)},y_{i(1)}),\ldots, (x_{i(d_1+d_2)},y_{i(d_1+d_2)}) } \subset \mathcal{W}_1\oplus \mathcal{W}_2$, and $(b)$ follows from the condition ${(\Gamma 1)}$. 
Since $\rank T_x \leq d_1$ follows from its definition and the dimension of $\ker T_y$ is $d_1$,
the condition $\rank T_x=d_1$ is a necessary condition for $\ker T_x \cap  \ker T_y = \{\zerom{d_1+d_2}{1}\}$.
Therefore, we have
\begin{align}
  &\Pr\bparen*{\ker T_x \cap  \ker T_y = \{\zerom{d_1+d_2}{1}\}\mid {(\Gamma 2')}\cap{(\Gamma 1)}}    \nonumber \\
= &\Pr\bparen*{\ker T_x \cap  \ker T_y \mid \rank T_x = d_1 \cap {(\Gamma 2')}\cap{(\Gamma 1)}} 
    \cdot \Pr\bparen*{\rank T_x = d_1 \mid{(\Gamma 2')}\cap{(\Gamma 1)}}. \label{eq:probab_221}
\end{align}
By applying Lemma \ref{lemm:space} for 
$(d_a,d_b,d_c,\mathcal{W}) := (d_1+d_2, d_1\!=\!\dim\ker T_y, d_2\!=\!\dim\ker T_x, \ker T_y)$, the first multiplicand of \eqref{eq:probab_221} equals to $1-O(1/{q'}^{-1})$.
From 
$\Pr\bparen*{\rank T_x=d_1\mid {(\Gamma 2')}\cap{(\Gamma 1)}} \geq \Pr\big[ \rank [t_1,\ldots, t_{d_1+d_2} ] = d_1 \mid t_1,\ldots, t_{d_1+d_2} \in \mathbb{F}_{q'}^{d_1} \big]$
and
Lemma \ref{lemm:space2},
the second multiplicand of \eqref{eq:probab_221} is greater than or equal to $1-O(1/{q'}^{-1})$.
Therefore, $\Pr\bparen*{{(\Gamma 2)}|{(\Gamma 2')}\cap{(\Gamma 1)} }  \geq 1-O(1/{q'}^{-1})$.

In summary,
we obtain
\begin{align*}
 &\Pr[{(\Gamma 1)}\cap{(\Gamma 2)}\cap{(\Gamma 3)}] = \Pr[{(\Gamma 1)}] \cdot \Pr[{(\Gamma 2')}] \cdot \Pr[{(\Gamma 2)}|{(\Gamma 2')}\cap{(\Gamma 1)}]  \\
\geq & \paren*{1-O\paren*{\frac{1}{q'}}}\cdot \paren*{ 1-\frac{n'^{m_i}}{{q'}^{m_i-a_i}}}\cdot \paren*{1-O\paren*{\frac{1}{q'}}} = 1-O\paren*{\max \Big\{\frac{1}{q'} , \frac{(n')^{m_i}}{(q')^{m_i-a_i}} \Big\}}.
\end{align*}

\subsection{Phase Error Probability} \label{subsec:phase}
In this subsection, we show that the original message $M'$ in the phase basis is correctly recovered
with probability at least $1-O\paren*{\max \Big\{\frac{1}{q'} , \frac{(n')^{m_i}}{(q')^{m_i-a_i'}} \Big\}}$.

\noindent{\it Step 1:\quad}
We derive a necessary condition for correct decoding of the original message $M'$ in phase basis. 
For the analysis of the phase error probability, we apply the same discussion as the bit error probability.
When a phase basis state $\phasek{M'}\in\cH'_{\mathrm{code}}$ is the input state of sender $S_i$, the encoded state is written as 
\begin{align}
\mathcal{E}_i^{SR_i,R_i}(\phasek{M'}) =
\sum_{\bar{E}_1' \in \mathbb{F}_{q'}^{m_i\times m_i},
\bar{E}_2' \in \mathbb{F}_{q'}^{a_i\times (n'-2m_i)}}
\phasek*{ \ph{U_{i,1}}
    \left[\begin{array}{ccc}
    \multicolumn{1}{c}{\multirow{3}{*}{$\bar{E}_1'$}} &  \multicolumn{1}{c}{\multirow{2}{*}{$R_{i,2}$}}  & \bar{E}_2' \\
    \multicolumn{1}{c}{} & \multicolumn{1}{c}{} & M' \\
    \multicolumn{1}{c}{} & \zerom{a'_i}{\mm}  & \zerom{a'_i}{n'-2\mm}\\
    \end{array}\right]
    \ph{U_{i,2}^{V_i}}   }, \nonumber
\end{align}
where we ignore normalizing factors and phase factors.

Since phase basis measurement and decoding operation $\mathcal{D}_i^{SR_i}$ commutes,
we first apply phase basis measurement, and then decode the measurement outcome for the analysis of the phase error probability.
Then, the phase basis measurement outcome $Y'$ on the network output of $T_i$ is written as 
\begin{align}
Y' = \ph{K}_{i,i} 
    \ph{U_{i,1} }
    \left[\begin{array}{ccc}
    \multicolumn{1}{c}{\multirow{3}{*}{$\bar{E}_1'$}} &  \multicolumn{1}{c}{\multirow{2}{*}{$R_{i,2}$}}  & \bar{E}_2' \\
    \multicolumn{1}{c}{} & \multicolumn{1}{c}{} & M' \\
    \multicolumn{1}{c}{} & \zerom{a'_i}{\mm}  & \zerom{a'_i}{n'-2\mm}\\
    \end{array}\right]
    \ph{U_{i,2}^{V_i}}
  + \ph{K}_{i^c} Z,  \nonumber
\end{align}
where 
$\bar{E}_1' \in \mathbb{F}_{q'}^{m_i\times m_i}$,
$\bar{E}_2' \in \mathbb{F}_{q'}^{a_i\times (n'-2m_i)}$
and 
$Z \in \mathbb{F}_{q'}^{(m-m_i)\times n'}$.
By Step D1, $Y'$ is decoded to 
\begin{align*}
\bar{Y}' =
Y'(\ph{U_{i,2}^{V_i}})^{-1} = 
\ph{K}_{i,i} \ph{U_{i,1} }
    \left[\begin{array}{ccc}
    \multicolumn{1}{c}{\multirow{3}{*}{$\bar{E}_1'$}} &  \multicolumn{1}{c}{\multirow{2}{*}{$R_{i,2}$}}  & \bar{E}_2' \\
    \multicolumn{1}{c}{} & \multicolumn{1}{c}{} & M' \\
    \multicolumn{1}{c}{} & \zerom{a'_i}{\mm}  & \zerom{a'_i}{n'-2\mm}\\
    \end{array}\right]
    + \ph{K}_{i^c} Z (\ph{U_{i,2}^{V_i}})^{-1} . \nonumber
\end{align*}
By Step D2, the measurement outcome $O_{i,2}$ is given as 
$O_{i,2} = \bar{Y}'^{\mathcal{B}} = 
\ph{K}_{i,i} \ph{U_{i,1} }
    \begin{bmatrix}
    \multicolumn{1}{c}{\multirow{2}{*}{$R_{i,2}$}}\\
    \multicolumn{1}{c}{} \\
    \zerom{a_i'}{\mm}
    \end{bmatrix}
    +
    (\ph{K}_{i^c} Z (\ph{U_{i,2}^{V_i}})^{-1})^{\mathcal{B}},
$
and $D_{i,2}^{R_{\!i,\!2},O_{\!i,\!2}}$ is found by Gaussian elimination to the right equation of \eqref{eq:equation} which is written as
\begin{align}
P_{\mathcal{W}_{i,2}} D_{i,2}^{R_{\!i,\!2},O_{\!i,\!2}} O_{i,2} &=
P_{\mathcal{W}_{i,2}} D_{i,2}^{R_{\!i,\!2},O_{\!i,\!2}}
    \paren*{
    \ph{K}_{i,i} \ph{U_{i,1} }
    \begin{bmatrix}
    \multicolumn{1}{c}{\multirow{2}{*}{$R_{i,2}$}}\\
    \multicolumn{1}{c}{} \\
    \zerom{a_i'}{\mm}
    \end{bmatrix}
    +
    (\ph{K}_{i^c} Z (\ph{U_{i,2}^{V_i}})^{-1})^{\mathcal{B}}}
=
\begin{bmatrix}
    \multicolumn{1}{c}{\multirow{2}{*}{$R_{i,2}$}}\\
    \multicolumn{1}{c}{} \\
    \zerom{a_i'}{\mm} \end{bmatrix}. \label{eq:Gaussian_elim_p}
\end{align}
Thus, the correct estimate of $M'$ is obtained when the following relation holds for $D_{i,2}^{R_{\!i,\!2},O_{\!i,\!2}}$ derived in \eqref{eq:Gaussian_elim_p}:
\begin{align}
P_{\mathcal{W}_{\!i\!,2}}\! D_{i,\!2}^{R_{i\!,2},O_{i\!,2}} 
\bar{Y}'^{\mathcal{C}}
 =
P_{\mathcal{W}_{i\!,2}} \!D_{i,\!2}^{R_{i\!,2},O_{i\!,2}} 
 \paren*{
    \ph{K}_{i\!,i} \ph{U_{i,\!1} }
    \begin{bmatrix}
    \bar{E}_2'\\
    M' \\
    \!\zerom{a'_i}{n'\!-\!2\mm}\!
    \end{bmatrix}
    +
    (\ph{K}_{i^c} Z (\ph{U_{i\!,2}^{V_i}})^{-1})^{\mathcal{C}}}
=
    \begin{bmatrix}
    \bar{E}_2'\\
    M' \\
    \!\zerom{a'_i}{n'\!-\!2\mm}\!
    \end{bmatrix}. \label{eq:prob_bit_p}
\end{align}

\noindent{\it Step 2:\quad}
In the next step, we show that the equation \eqref{eq:prob_bit_p} holds with probability at least
$1-O\paren*{\max \Big\{\frac{1}{q'} , \frac{(n')^{m_i}}{(q')^{m_i-a'_i}} \Big\}}$,
which shows the desired statement by combining Step 1.

In the same way as Subsection \ref{subsec:bit}, 
the conditions ${(\Gamma 1)}$, ${(\Gamma 2)}$ and ${(\Gamma 3)}$ of
Lemma \ref{lemm:prob_bit} in the following case imply Eq. \eqref{eq:prob_bit_p};
\begin{gather*}
\mathcal{W}_1:=\col\paren*{\ph{K}_{i,i} \ph{U_{i,1}}
    \begin{bmatrix}
    \multicolumn{1}{c}{\multirow{2}{*}{$R_{i,2}$}}\\
    \multicolumn{1}{c}{} \\
    \zerom{a_i'}{\mm}
    \end{bmatrix}}, \quad
\mathcal{W}_2:=\col\paren*{ \ph{K}_{i^c} Z (\ph{U_{i,2}^{V_i}})^{-1}}, 
\quad \mathcal{W}_1':= \mathcal{W}_{i,2}, 
\quad \xm:=m_i,\\
[x_1,\ldots, x_{\xm}] := \ph{K}_{i,i} \ph{U_{i,1}}
    \begin{bmatrix}
    \multicolumn{1}{c}{\multirow{2}{*}{$R_{i,2}$}}\\
    \multicolumn{1}{c}{} \\
    \zerom{a_i'}{\mm}
    \end{bmatrix}, \quad
[y_1,\ldots, y_{\xm}] := (\ph{K}_{i^c} Z (\ph{U_{i,2}^{V_i}})^{-1})^{\mathcal{B}}, \\
[r_1,\ldots, r_{\xm}] := 
    \begin{bmatrix}
    \multicolumn{1}{c}{\multirow{2}{*}{$R_{i,2}$}}\\
    \multicolumn{1}{c}{} \\
    \zerom{a_i'}{\mm}
    \end{bmatrix},\quad
     A := (\ph{K}_{i,i} \ph{U_{i,1}})|_{\mathcal{W}_1'},
\quad (d_0,d_1,d_2) := (m_i, m_i-a'_i, \rank \ph{K}_{i^c} Z), 
\end{gather*}
where $\mathcal{W}_{i,2}$ is defined in Step D2 of Subsection \ref{subsec:dec}.
Also, in the same way,
the conditions ${(\Gamma 1)}$, ${(\Gamma 2)}$ and ${(\Gamma 3)}$ holds
with probability at least 
$1-O\paren*{\max \Big\{\frac{1}{q'} , \frac{(n')^{m_i}}{(q')^{m_i-a'_i}} \Big\}}$.

\section{Code Construction Without Free Classical Communication}   \label{sec:without_sr}

We show that our code in Theorem \ref{theo:main1} can be implemented without the assumption of negligible rate shared randomness.
The paper \cite{YSJL14} shows the following Proposition \ref{prop:secret_classical} by 
constructing a secret and correctable classical communication protocol for the classical unicast linear network.
Due to the relation between the phase error and the information leakage in the bit basis \cite[Lemma 5.9]{Haya3},
we find that the dimension of leaked information is $a_i'$ 
in the information transmission from the sender $S_i$ to the receiver $T_i$.
We apply Proposition \ref{prop:secret_classical} to the sender-receiver pair $(S_i, T_i)$
with $c_1:=a_i$ and $c_2:=a_i'$.
Therefore, the protocol of Proposition \ref{prop:secret_classical} can be implemented in our multiple-unicast network by preparing 
the input state of $S_i$ in the bit basis.
By attaching Proposition \ref{prop:secret_classical} to our code in the above method, 
we can implement our code satisfying Theorem \ref{theo:main2}.

\begin{prop}[{\cite[Theorem 1]{YSJL14}}]
Let $q_1$ be the size of the finite field which is the information unit of the network edges.
We assume the inequality $c_1 +c_2 < c_0$ for the classical network where 
$c_0$ is the transmission rate from the sender $S$ to the receiver $T$,
$c_1$ is the rate of noise injection, and
$c_2$ is the rate of information leakage. 
Define $q_2:=q_1^{c_0}$.
Then, there exists a $k$-bit transmission protocol of block-length $n_1:=c_0(c_0-c_2+1)k$ over $\mathbb{F}_{q_2}$ such that
$P_{err} \leq kc_0/q_2 \text{  and  }
I(M;E) = 0$,
where $P_{err}$ is the error probability and $I(M;E)$ is the mutual information between the message $M\in\mathbb{F}_2^k$ and the leaked information $E$.
\label{prop:secret_classical}
\end{prop}

The proof of Theorem \ref{theo:main2} takes a similar method to the proof of \cite[Theorem 3.2]{SM18}.
\begin{IEEEproof}[Proof of Theorem \ref{theo:main2}]
To construct the code satisfying the conditions of Theorem \ref{theo:main2}, we generate the shared randomness $SR_i$ by Proposition \ref{prop:secret_classical}
and then apply the code in Section \ref{sec:code}.
To apply Proposition \ref{prop:secret_classical} in our quantum network,
we prepare the input state as a bit basis state. 
Given a block-length $n$, we take $q_1 = q^{\beta}$ such that 
$\beta = \lfloor \frac{2\log_2 \log_2 n }{m_i\log_2 q} \rfloor$ i.e., $q_2/ (\log n)^2=q_1^{m_i}/ (\log n)^2\to 1$,
and $q'=q^\alpha$ such that 
$\alpha=\lfloor \frac{(m_i+2)\log_2 n}{\log_2 q} \rfloor$ i.e., $q'/n^{m_i+2}\to 1$.

First, by the protocol of Proposition \ref{prop:secret_classical} with $(c_0,c_1,c_2):=(m_i,a_i,a_i')$, the sender $S_i$ and the receiver $T_i$ share the randomness $SR_i$.
Since $SR_i$ consists of $m_i(2m_i-a_i-a'_i+4)$ elements of $\mathbb{F}_{q'}$,
 the number of bits to be shared is 
 \begin{align*}
 k = \ceil*{ m_i(2m_i-a_i-a'_i+4)\log_2 q'} &= \ceil*{ m_i(2m_i-a_i-a'_i+4)\floor*{\frac{(m_i+2)\log_2 n}{\log_2 q}}\log_2 q} \\
 &\leq \ceil*{ m_i(m_i+2)(2m_i-a_i-a'_i+4)\log_2 n}.
 \end{align*}
The error probability is $P_{err} \!\leq (m_i/q_1^{m_i})\cdot \ceil*{ m_i(m_i\!+\!2)(2m_i\!-\!a_i\!-\!a'_i\!+\!4)\log_2 n } \! =\! O\paren*{\!\frac{\log_2 n }{(\log_2 n)^2}\!} \!\to 0,$
and the block-length over $\mathbb{F}_q$ is 
 $$n_1 \!=\! m_i(m_i\!-\!a_i'\!+\!1)k\beta \!\leq\! m_i(m_i\!-\!a_i'\!+\!1)\cdot \ceil*{ m_i(m_i\!+\!2)(2m_i\!-\!a_i\!-\!a'_i\!+\!4)\log_2 n} \cdot \floor*{ \frac{2\log_2 \log_2 n }{m_i\log_2 q}},$$ 
which implies $n_1/n \to 0$. Therefore, the sharing protocol is implemented with negligible rate uses of the network.

Next, we apply the code in Section \ref{sec:code} with the extended field of size $q'$ and $n_2:=n-n_1$ uses of the network.
The relation $n_2/n = (n-n_1)/n \to 1$ holds and therefore the field size $q'$ satisfies 
 $n_2\cdot (n_2')^{m_i}/(q')^{m_i-\max\{a_i,a_i\}}\to 0$ where $n_2':=n_2/\alpha.$
Thus, this code implements the code in Theorem \ref{theo:main2}. 
\end{IEEEproof}

\section{Examples of Network}   \label{sec:network_examples}
In this section, we give several network examples that our code can be applied.

First, as the most trivial case, if $\rank K_{i,i}=m_i$ and any distinct sender-receiver pairs do not interfere with each other, i.e,  $K_{i,j}$ ($i\neq j$) are zero matrices, the network operation $K$ is a block matrix.
This is the case where each pair independently communicates.
In this case, our code is implemented with the rate $m_i$.

\subsection{Simple Network in Fig. \ref{fig:butterfly}} \label{subsec:betterfly}

In the network in Fig. \ref{fig:butterfly}, the network and node operations are described as 
\begin{align*}
K =  \begin{bmatrix} 1 & 0 & 0 & 0 \\0 & 1 & 1 & 0\\ 0 & 0 & 1& 0 \\ 0 & 0 & 0 & 1 \end{bmatrix}, \quad
\ph{K} =  \begin{bmatrix} 1 & 0 & 0 & 0 \\0 & 1 & 0 & 0\\ 0 & -1 & 1& 0 \\ 0 & 0 & 0 & 1 \end{bmatrix}, \quad
A_1 =  \begin{bmatrix} 1 & 1 \\ 0 & 1 \end{bmatrix}.
\end{align*}
When we consider the transmission from $S_1$ to $T_1$,
the rates of bit and phase interferences are
\begin{align*}
\rank K_{1^c} = \rank \begin{bmatrix} 0 & 0 \\ 1 & 0 \end{bmatrix} = 1, \quad
\rank \ph{K}_{1^c} = \rank \begin{bmatrix} 0 & 0 \\ 0 & 0 \end{bmatrix} = 0.
\end{align*}
In this network, by constructing our code with $(m_1,a_1, a'_1):=(2,1,0)$, our coding protocol transmits the state of rate $m_1-a_1-a_1'=1$ asymptotically from $S_1$ to $T_1$.

\subsection{Network with Bit Interference from One Sender}

As a generalization of the network in Fig. \ref{fig:butterfly},
consider the case where the network consists of two sender-receiver pairs, and there is no bit interference from the sender $S_1$ to receiver $T_2$.
The network operation of this network can be described by $\lefto{K}$ with 
\begin{align*}
K = 
\begin{bmatrix}
K_{1,1} & K_{1,2}   \\
\zerom{m_2}{m_1} & K_{2,2}
\end{bmatrix}, \quad
\ph{K} = 
\begin{bmatrix}
(K_{1,1}^{\mathrm{T}})^{-1} & \zerom{m_1}{m_2}   \\
-(K_{2,2}^{\mathrm{T}})^{-1}K_{1,2}^{\mathrm{T}}(K_{1,1}^{\mathrm{T}})^{-1} & (K_{2,2}^{\mathrm{T}})^{-1}
\end{bmatrix}
.
\end{align*}
In this network, there is no phase interference from the sender $S_2$ to receiver $T_1$, and
the other two rates $\rank K_{1,2}$ and $\rank (K_{2,2}^{\mathrm{T}})^{-1}K_{1,2}^{\mathrm{T}}(K_{1,1}^{\mathrm{T}})^{-1}$ coincide from
 $\rank K_{1,2} =\rank K_{1,2}^{\mathrm{T}}  = \rank (K_{2,2}^{\mathrm{T}})^{-1}K_{1,2}^{\mathrm{T}}(K_{1,1}^{\mathrm{T}})^{-1}$.
Therefore, by implementing our code with $a_i,a'_i$ ($i=1,2$) satisfying $\rank K_{1,2} \leq a_1, a'_2< m_i$ and $a_1'=a_2 := 0$,
each sender-receiver pair can transmit the states.

Moreover, we generalize the above network for arbitrary $\p$ sender-receiver pairs where the interferences are generated only from one sender $S_1$.
In this network, the network operation is given by the unitary operator $\lefto{K}$ with $K$ defined as follows:
\begin{align*}
&
\begingroup
\renewcommand*{\arraystretch}{0.8}
\setlength\arraycolsep{2pt}
K =
\begin{bmatrix}
K_{1,1} & K_{1,2}  &  K_{1,3}                   &\cdots &  K_{1,\p} \\
\zerom{m_2}{m_1}& K_{2,2} & \zerom{m_2}{m_3}    &\cdots &  \zerom{m_2}{m_\p}\\
\vdots          & \vdots  & \vdots              &\ddots &  \vdots  \\
\zerom{m_\p}{m_1}& \zerom{m_\p}{m_2} &  \zerom{m_\p}{m_3}    &\cdots &  K_{\p,\p} \\
\end{bmatrix},
\endgroup
\enskip
\begingroup
\renewcommand*{\arraystretch}{0.8}
\setlength\arraycolsep{2pt}
\ph{K} = 
\begin{bmatrix}
(K_{1,1}^{\mathrm{T}})^{-1} & \zerom{m_1}{m_2}  & \zerom{m_1}{m_3}  &  \cdots  & \zerom{m_1}{m_\p} \\
-(K_{2,2}^{\mathrm{T}})^{-1}K_{1,2}^{\mathrm{T}}(K_{1,1}^{\mathrm{T}})^{-1} & (K_{2,2}^{\mathrm{T}})^{-1} & \zerom{m_2}{m_3} & \cdots     &  \zerom{m_1}{m_\p}\\
\vdots & \vdots & \vdots &\ddots & \vdots  \\
-(K_{\p,\p}^{\mathrm{T}})^{-1}K_{1,\p}^{\mathrm{T}}(K_{1,1}^{\mathrm{T}})^{-1} & \zerom{m_\p}{m_2} &  \zerom{m_\p}{m_3}     &   \cdots   & (K_{\p,\p}^{\mathrm{T}})^{-1}  \\
\end{bmatrix},
\endgroup
\end{align*}
where the ranks of $m_i\times m_i$ matrices $K_{i,i}$ are $m_i$, resepctively. 
In this network, if $a_i, a_i'$ ($i=1,\ldots,r$) are set to 
$a_1 \geq \rank [K_{1,2} \pp K_{1,3} \pp \cdots \pp K_{1,\p}]$,
$a'_i \geq \rank K_{1,i} $ ($i=2,\ldots,\p$),
and  $a_1'=a_2=a_3 = \cdots = a_\p \geq 0$
and the condition $a_i+a'_i < m_i$ holds,
the sender $S_i$ can send to the receiver $T_i$ the rate $m_i-a_i-a'_i$ state asymptotically by our code.

\subsection{Network with Two Way Bit Interferences}
In this subsection, we consider the code implementation over the network described as follows:
The size $q$ is $3$, 
there exists two pairs $(S_1,T_1)$ and $(S_2,T_2)$ in the network,
$S_1,S_2,T_1,T_2$ are connected to three edges,
and the network operation is given by $\lefto{K}$ of
\begin{align*}
\begingroup
\renewcommand*{\arraystretch}{0.9}
\setlength\arraycolsep{3pt}
K = \begin{bmatrix}
K_{1,1} & K_{1,2}   \\
K_{2,1} & K_{2,2}
\end{bmatrix}
=\begin{bmatrix}
1 & 0 & 0 & 1 & 0 & 0 \\
0 & 1 & 0 & 0 & 0 & 0 \\
0 & 0 & 1 & 0 & 0 & 0 \\
-1& 0 & 0 & 1 & 0 & 0 \\
0 & 0 & 0 & 0 & 1 & 0 \\
0 & 0 & 0 & 0 & 0 & 1 \\
\end{bmatrix},\quad
\ph{K} = \begin{bmatrix}
2 & 0 & 0 & 2 & 0 & 0 \\
0 & 1 & 0 & 0 & 0 & 0 \\
0 & 0 & 1 & 0 & 0 & 0 \\
-2& 0 & 0 & 2 & 0 & 0 \\
0 & 0 & 0 & 0 & 1 & 0 \\
0 & 0 & 0 & 0 & 0 & 1 \\
\end{bmatrix}.
\endgroup
\end{align*}
Then, differently from the previous examples, there are bit interferences both from $S_1$ to $T_2$ and from $S_2$ to $T_1$ because $K_{1,2}$ and $K_{2,1}$ are not zero matrix.

In the above network, we construct our code for $S_1$ to $T_1$ with $(m_1,a_1,a_1'):=(3,1,1)$.
Then, our code implements the rate $m_i-a_i-a_i'=3-1-1=1$ quantum communication asymptotically from the relations
\begin{align*}
\rank K_{11} \!=\! 
\rank \ph{K}_{11} \!=\!   m_1 \!=\! 3, \!\!\quad
\rank K_{1^c} \!=\! &
\rank
\begingroup
\renewcommand*{\arraystretch}{0.8}
\setlength\arraycolsep{2pt}
\begin{bmatrix}
1 & 0 & 0\\
0 & 0 & 0\\
0 & 0 & 0
\end{bmatrix} \!=\! 1,
 \!\!\quad
\rank \ph{K}_{1^c} \!=\! 
\rank
\begin{bmatrix}
2 & 0 & 0\\
0 & 0 & 0\\
0 & 0 & 0
\end{bmatrix} \!=\! 1.
\endgroup
\vspace{-5mm}
\end{align*}

\section{Conclusion} \label{sec:conclusion}
In this paper, we have proposed a quantum network code for the multiple-unicast network with quantum invertible linear operations.
As the constraints of information rates, 
we assumed that the bit and phase transmission rates from $S_i$ to $T_i$ without interference are $m_i$ ($m_i=\rank K_{i,i}=\rank \ph{K}_{i,i}$),
the upper bounds of the bit and phase interferences are $a_i, a_i'$, respectively ($\rank K_{i^c} \leq a_i$, $\rank \ph{K}_{i^c}\leq a'_i$),
and $a_i+a_i'<m_i$ holds.
Under these constraints, our code achieves the rate $m_i-a_i-a_i'$ quantum communication by asymptotic $n$-use of the network.
The negligible rate shared randomness plays a crucial role in our code, and it is realized by attaching the protocol in \cite{YSJL14}.

Our code can be applied for the multiple-unicast network with the malicious adversary. 
When the eavesdropper attacks 
at most $a''_i$ edges connected with the sender $S_i$ and the receiver $T_i$,
if $a_i+a'_i+2a''_i < m_i$ holds, our code implements the rate $m_i-a_i-a'_i-2a''_i$ quantum communications asymptotically.
This fact can be shown by integrating the methods in this paper and \cite{SM18}.

\end{document}